\UseRawInputEncoding
\documentclass[reprint,
 amsmath,amssymb,
 aps,
]{revtex4-2}

\usepackage{graphicx}
\usepackage{dcolumn}
\usepackage{bm}

\usepackage{subcaption}
\usepackage{xcolor}
\usepackage{caption}

\begin{document}

\preprint{APS/123-QED}

\title{Two-step estimators of high dimensional correlation matrices}

\author{Andr\'es Garc\'ia-Medina}
\email{andres.garcia@cimat.mx}
 \affiliation{Centro de Investigaci\'on en Matem\'aticas, Unidad Monterrey, Av. Alianza Centro 502, PIIT 66628, Apodaca, Nuevo Le\'on, M\'exico}%
 \affiliation{Consejo Nacional de Ciencia y Tecnolog\'ia, Av. Insurgentes Sur 1582, Col. Cr\'edito Constructor 03940, Ciudad de M\'exico, M\'exico}
\author{Salvatore Miccich\`e}%
\email{salvatore.micciche@unipa.it}
\affiliation{Dipartimento di Fisica e Chimica Emilio Segr\`e, Universit\`a
degli Studi di Palermo, Viale delle Scienze, Ed. 18, 90128, Palermo, Italy}%

\author{Rosario N. Mantegna}
\email{rosario.mantegna@unipa.it}
\affiliation{Dipartimento di Fisica e Chimica Emilio Segr\`e, Universit\`a
degli Studi di Palermo, Viale delle Scienze, Ed. 18, 90128, Palermo, Italy}
\affiliation{Complexity Science Hub Vienna, Josefstdter Strasse 39, 1080 Vienna, Austria}

\date{\today}

\begin{abstract}
We investigate block diagonal and hierarchical nested stochastic multivariate Gaussian models by studying their sample cross-correlation matrix on high dimensions. By performing numerical simulations, we compare a filtered sample cross-correlation with the population cross-correlation matrices by using several rotationally invariant estimators (RIE) and hierarchical clustering estimators (HCE) under several loss functions. We show that at large but finite sample size, sample cross-correlation filtered by RIE estimators are often outperformed by HCE estimators for several of the loss functions. We also show that for block models and for hierarchically nested block models the best determination of the filtered sample cross-correlation is achieved by introducing two-step estimators combining state-of-the-art non-linear shrinkage models with hierarchical clustering estimators.
\end{abstract}

\maketitle


\section{Introduction}

In recent years, many research areas have dealt with multivariate time series. Examples are physics, neuroscience, finance, climatology, genomics, etc. In all these research areas, investigators perform $n$ measurements of a system characterized by $p$ variables, obtaining an observation matrix {\bf Y} of dimension $p \times n$. After standardizing the $p$ series of $n$ records, one can compute the $p \times p$ sample cross-correlation matrix {\bf E}. Sample cross-correlation matrices computed from a finite set of multivariate data generally differ from the population cross-correlation matrix {\bf C} associated with the model generating multivariate data. Since the seminal work of Marcenko and Pastur \cite{marvcenko1967distribution}, many studies have considered the spectral properties of sample cross-correlation matrices and have used these theoretical results to set up a null model useful to discriminate information that can be extracted from data, i.e., information not compatible with a null model, and information hard to be distinguished from noise, i. e., a null model, in empirical data \cite{bun2017cleaning}.   

Comparing sample and population cross-correlation requires choosing a loss function, i.e., a function specifying a penalty for an incorrect estimate from the underlying statistical model. In the literature, several loss functions have been proposed, and the choice of a specific one must be related to the specific problem considered. The most used loss functions are Frobenius loss, Stein loss, and Kullback-Leibler divergence.

In~\cite{tumminello2007kullback,tumminello2007shrinkage}, the authors analytically demonstrate that the expected Kullback-Leibler~(KL) divergence of a sample correlation matrix concerning the population matrix does not depend on the reference model. The authors used the KL divergence to measure how informative the filtered correlation matrices are when applying spectral and hierarchical clustering techniques separately. The studies perform simulations against factor model structures \cite{tumminello2007hierarchically} and empirical studies with financial time series listed on US equity markets. As spectral techniques, the authors use two variations of methods known in the literature as the \emph{clipping} technique. The clipping~(also known as filtering or denoising in the econophysics community) technique was initially proposed in the work~\cite{laloux1999noise,plerou2002random} and later cataloged in the family of Rotationally Invariant Estimators~(RIE) by ~\cite{bun2017cleaning}. In particular, the clipping technique is associated with the spiked covariance matrix model~\cite{johnstone2001distribution}.

RIE models for estimating the covariance matrix have been known in the mathematical statistics community since Stein ~\cite{stein1975estimation} proposed them under the name \emph{rotation-equivariant} estimators. His idea was to keep the eigenvectors of the sample covariance matrix while shrinking its eigenvalues. They were proposed in the classical paradigm when the number of observations is much greater than the number of variables. Ledoit and Wolf have been promoting of these methods on the high-dimensional stage. In~\cite{ledoit2004well}, they proposed an optimal linear shrinkage using Random Matrix Theory~(RMT) concepts. Later, in~\cite{ledoit2011eigenvectors}, the first non-linear shrinkage model based on RMT and asymptotic theory is proposed. Their numerical implementation is given in~\cite{ledoit2017numerical}.
On the other hand, Bun, Bouchaud, and Potters~\cite{bun2017cleaning} suggest a different numerical approach that is easier to implement. Both are approximations of the same model. Finally, Ledoit and Wolf give a kernel-based solution that is essentially analytical and drastically improves the computation time by two orders of magnitude~\cite{ledoit2020analytical}. This solution is valid for general correlation structures but does not consider autocorrelations.
Burda and Jarosz tackle the autocorrelation structure in a recent work~\cite{Burda2022} by using concepts of RMT and free probability.

The previous formulations lead to considering non-linear shrinkage formulas to estimate the population eigenvalues from the empirical ones and reconstruct the correlation matrices using the empirical eigenvectors. As such, they belong to the RIE family of estimators. These non-linear shrinkage formulas are optimal with the Frobenius loss function. The problem of applying different loss functions is of current interest, as stated, for example, in~\cite{bun2017cleaning,donoho2018optimal}, where authors propose to quantify the information kept by the optimal RIE compared to several estimators and metrics.

Several empirical covariance matrices present a spectrum compatible with the so-called 
spiked covariance matrix model~\cite{johnstone2001distribution,donoho2018optimal}, i.e., covariance matrices with an eigenvalue spectrum characterized by a few number of large isolated eigenvalues distinct from bulk eigenvalues. So-called hierarchically nested factor models, i.e., factor models with nested factors affecting distinct subgroups of elements of the systems present a spiked eigenvalue spectrum \cite{lillo2005spectral,tumminello2007spectral}. One of the main results of spiked covariance models is the presence of top eigenvector inconsistency \cite{donoho2018optimal}, i.e., the observation that the angle between sample eigenvectors and the corresponding population eigenvectors have non zero limits. This implies that the optimal choice of the nonlinear shrinkage function of eigenvalues might depend significantly by the specific loss function chosen. Top eigenvector inconsistency also suggests that filtering by rotationally invariant estimators using sample eigenvectors might miss some aspects of the population matrix. Another limitation of RIE methods concerns sample eigenvectors associated to small eigenvalues. They usually comprise components covering the entire set of elements therefore presenting an eigenvector orientation quite distinct from localized eigenvectors associated with a  correlated dynamics of a small group of elements.     

The above observations have motivated an alternative filtering procedure of spiked correlation matrices based on hierarchical clustering. In fact, in standardized random multivariate variables with correlation matrices characterized by positive correlation coefficients there is a one to one correspondence between the cophenetic matrix of a hierarchical clustering and a hierarchically nested factor model \cite{tumminello2007hierarchically}. A hierarchical clustering procedure therefore provides a correlation matrix equivalent to a hierarchically nested factor model \cite{tumminello2010correlation}. The effectiveness of filtering of a correlation matrix by hierarchical clustering has been documented in several studies primarily associated with the problem of portfolio optimization in finance 
\cite{tola2008cluster,pantaleo2011improved,bongiorno2021cleaning,bongiorno2021covariance,bongiorno2022reactive}.

Although the subject of RIE methods has been analytically studied extensively ~\cite{ledoit2021shrinkage}, some assumptions about eigenvectors might induce relevant limitations in the presence of complex systems characterized by correlation matrices with a hierarchically nested structure. Since no analytical results exist about optimal filtering by hierarchical clustering, we conduct a series of numerical experiments to evaluate the performances of different filtering methods based on RIE and on hierarchical clustering for different loss functions. Our numerical results suggest that RIE methods and hierarchical clustering methods give comparable results for systems whose population matrix is a spiked correlation matrix. We hope our results can stimulate the development of new analytical results both for RIE and HC filtering estimators.

Specifically, we numerically analyze block diagonal and hierarchical nested models on high dimensions and compare their behavior under several loss functions when applying RIE and hierarchical clustering estimators. We are also introducing two-step estimators that combine state-of-the-art non-linear shrinkage models with  hierarchical clustering estimators. These estimators outperform several of the most used estimators when the model of multivariate series is a block model or a hierarchically nested block model and when the statistical properties of records are Gaussian.

The paper is organized as follows. Section 2 describes the estimators proposed in this work. Section 3 introduces the loss functions used to evaluate the difference between filtered sample correlation and population correlation when applying each estimator. Section 4 gives the specifications of the models studied. Section 5 shows the main results obtained. Section 6 analyzes and discusses the findings found.

\section{Estimators}
\label{estimators}
For the sake of completeness, this section presents the estimators of the correlation matrix that we will use in our numerical analyses. These estimators can be grouped into three classes. The first ones belong to the RIE family, the second ones are of the hierarchical clustering type, and the third class combines both, which we denote as two-step estimators. It is important to emphasize that the first class of estimators is designed to deal with the estimation uncertainty inherent in the high-dimensional scenario when the number of variables is of the same order as the number of observations. The second class of estimators deals with the estimation uncertainty associated with the structure of the correlation blocks between variables. Therefore, it is focused on better detection of the financial sectors. Finally, the third class of estimators deals with both types of noise.

\subsection{Rotationally Invariant Estimators~(RIE)}

The RIE has the property that the sample correlation matrix $\mathbf{E}$ can be rotated by some orthogonal matrix $\mathbf{O}$ and its estimation, denoted as $\mathbf{\Xi}$, must be rotated in the same direction. 
Therefore, $\mathbf{\Xi}(\mathbf{E})$ can be diagonalized on the same basis as $\mathbf{E}$ except for a fixed rotation $\Omega$. In this way, $\mathbf{\Xi}(\mathbf{E})$ has the same eigenvectors as $\mathbf{E}$ and it is possible to write
\begin{equation}
\mathbf{\Xi}(\mathbf{E}) = \sum_{i=1}^{p} \xi_i v_i v_i',
\label{XI}
\end{equation}
where $v_i$ are the eigenvectors of $\mathbf{E}$, and $\xi_i$ is a function of the eigenvalues $[\lambda_j]_{j\in\{1,p\}}$ of $\mathbf{E}$.

The empirical correlation matrix $\mathbf{E}$ is a trivial example that satisfies this condition. Then, a \emph{naive estimator} is
\begin{equation}
    \mathbf{\Xi}^{naive} = \mathbf{E}
\end{equation} 

A classical RMT filter is proposed in~\cite{laloux1999noise,plerou2002random} and is expressed as
    \begin{equation}
     \xi^{RMT} = \left\{
                \begin{array}{ll}
		         \bar{\lambda}      & if \quad \lambda_k < (1 + \sqrt{q})^2\\
	        	 \lambda_k    & \text{otherwise}
                \end{array}
                \right.
                \label{RMT}
    \end{equation}
where $\bar{\lambda}$ represents the eigenvalues average below the Marchenko-Pastur law's upper bound. Then, the estimated correlation matrix is given by
    \begin{equation}
        \mathbf{\Xi}^{RMT} = \sum_{i=1}^{p} \xi_i^{RMT} v_i v_i',
    \end{equation}
  
A nonlinear shrinkage formula of the RIE family to estimate the unbiased covariance matrix when $\mathbf{C}$ has a general form given by~\cite{ledoit2011eigenvectors}
\begin{eqnarray}
\xi^{LP}_k  &=&  \lim_{\epsilon\rightarrow 0^{+}}\frac{\lambda_k}{|1-q+q\lambda_k G_E(\lambda_k-i\epsilon)|^2}\\
&=&\frac{\lambda_k}{|1+u_k|^2} \\
&=& \frac{\lambda_k}{(\alpha_k+1)^2 +\beta_k^2},
\label{LedoitPeche2011}    
\end{eqnarray}
where $\lambda_k$ is an eigenvalue of $\mathbf{E}$, $G_E$ is the Stieltjes transform of $\mathbf{E}$, and since we are close to the real axis, the Sokhotski-Plemelj formula applies
\begin{equation}
u_k = qT_E(\lambda_k -i0^{+}) = \alpha_k + i\beta_k, 
\end{equation}
where $T_E = zG_E(z)-1$, 
$\alpha_k = q(\pi\lambda_k h_E(\lambda_k)-1)$, and $\beta_k = q\pi \lambda_k\rho_E(\lambda_k)$. Here, $h_E$ denotes the Hilbert transform of $\mathbf{E}$ and $\rho_E$ its eigenvalue density.
The corresponding estimated correlation matrix is given by the following expression:
    \begin{equation}
        \mathbf{\Xi}^{LP} = \sum_{i=1}^{p} \xi_i^{LP} v_i v_i',
    \end{equation}

A recent proposal for non-linear shrinkage expression is due to Burda and Jarosz~\cite{Burda2022}, who incorporate autocorrelation into data-generating processes through a matrix $\mathbf{A}$. The authors gave explicit solutions for some specific models of the Vector Autoregressive Moving Average~(VARMA) family~\cite{lutkepohl2005new}:
\begin{equation}
    Y_{i,a} = \sum_{\beta=1}^{r_1} b_{\beta} Y_{i,a-\beta}+\sum_{\alpha=0}^{r_2}a_{\alpha} \epsilon_{i,a-\alpha}.
\end{equation}
The key element to analytically incorporate autocorrelations is the $\mathcal{S}$-transform of the associated matrix of coefficients $\mathbf{A}$, which in principle, is not trivial to compute. Calculating $\mathcal{S}$ requires some knowledge of the free probability~\cite{mingo2017free, potters2020first}. Burda and Jarosz explicitly solve the model for $(r_1,r_2)\in\{(1,1),(1,0),(2,0),(0,1),(0,2)\}$. 
The nonlinear shrinkage formula has the general form
    \begin{equation}
    \xi^{BJ}_k  =  \frac{\lambda_k Im\{1/Z_A(u_k)\}}{Im\{u_k\}}
    \label{BurdaJarosz2022}
    \end{equation}
where $Z_A$ is the Z-transform of $\mathbf{A}$.
Consequently, the optimal estimator in the Frobenius sense is:
    \begin{equation}
        \mathbf{\Xi}^{BJ} = \sum_{i=1}^{p} \xi_i^{BJ} v_i v_i',
    \end{equation}
Notice that when $\mathbf{A}=\mathbf{I}$, the $\mathcal{S}$-transform of $A$ is given by
\begin{equation}
S_A(t) = \frac{t+1}{tZ_A(t)} = 1 \Rightarrow Z_A(t) = \frac{t+1}{t}
\end{equation}
Then
\begin{equation}
\xi^{BJ}_k  =  \frac{\lambda_k Im\{1/Z_A(u_k)\}}{Im\{u_k\}} = \frac{\lambda_k}{(\alpha_k+1)^2+\beta_k^2},
\end{equation}
and we recover Eq.~\ref{LedoitPeche2011}.
    
In particular, the combination of parameters $a_0=\sqrt{1-b_1^2}, b_1=e^{-1/\tau}$, $a_{i}=b_{i-1} = 0$~(for $i>1$) represents the exponential decay model for which $Z_A$ is known analytically~\cite{burda2005spectral, burda2011applying}
\begin{equation}
Z_A(z) = \eta + \sqrt{\eta^2-1+1/z^2},
\label{Zeta}
\end{equation}
where $\eta=\coth(1/\tau)$.

A further estimator proposed from a data-driven approach employs the technique known as \emph{moving window cross-validation}~(mwcv) and is denoted as the oracle estimator~\cite{bartz2016cross}. It is important to mention that this estimator approximates the state-of-the-art nonlinear shrinkage~\cite{bun2018overlaps}. The expression to estimate the population eigenvalues is given by the expression:
\begin{equation}
    \xi_i^{mwcv} = \frac{1}{K} \sum_{\mu}^{K-1} \langle \lambda_i^{train,\mu} |\mathbf{E}^{^{test,\mu}} | \lambda_i^{train,\mu} \rangle,
    \label{mwcv}
\end{equation}
where $K=(T_{total}-T)/T_{out}$. The idea is to set $T$ observations as a train and $T_{out}$ as a test in a moving window scenario of the entire sample sequence of length $T_{total}=KT_{out}+T$. Here, $\lambda_i^{train,\mu}$ represents the eigenvalues of the training sample in window $\mu$ and $\mathbf{E}^{test,\mu}$ the test sample covariance matrix in window $\mu$.

    \begin{equation}
    \mathbf{\Xi}^{mwcv} = \sum_{i=1}^{p} \xi^{mwcv}_i v_i v_i',
    \end{equation}
    
\subsection{Hierarchical clustering estimators}

\begin{figure}
     \centering
     \begin{subfigure}[b]{0.3\textwidth}
         \centering
        \includegraphics[scale=0.45]{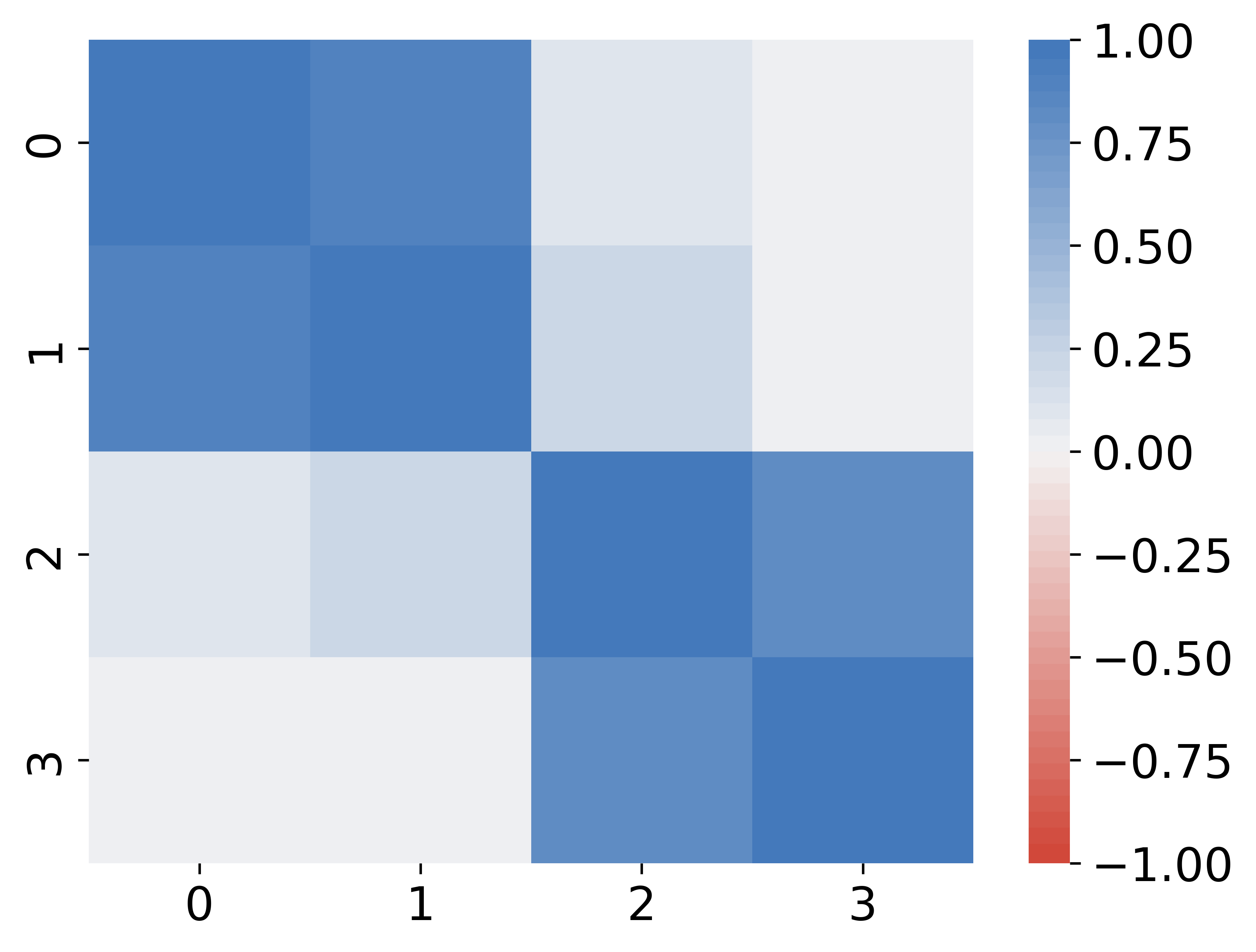}
         \caption{}
     \end{subfigure}
     \hfill
     \begin{subfigure}[b]{0.3\textwidth}
         \centering
\includegraphics[scale=0.25]{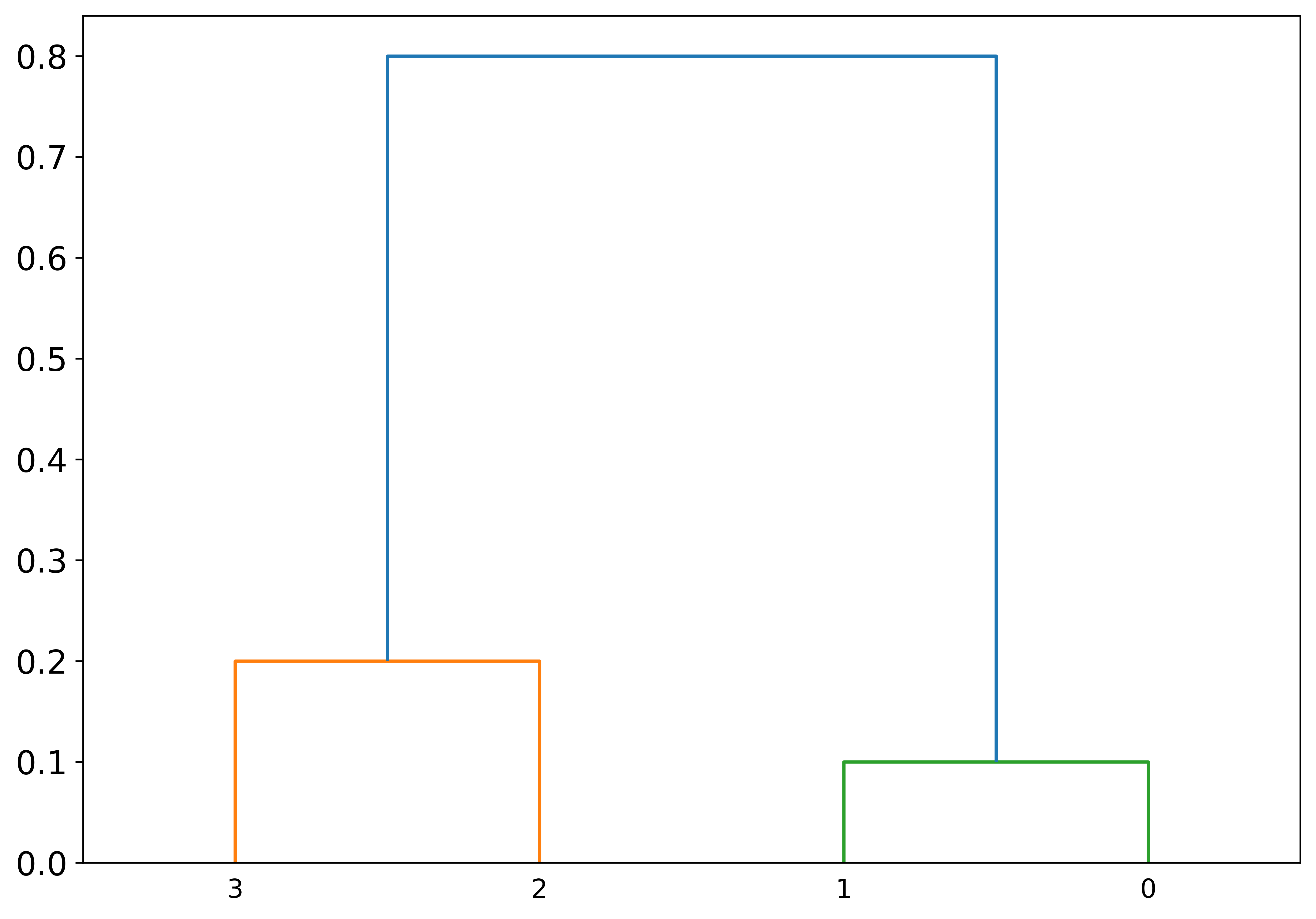}
         \caption{}
     \end{subfigure}
     \hfill
     \begin{subfigure}[b]{0.3\textwidth}
         \centering
\includegraphics[scale=0.45]{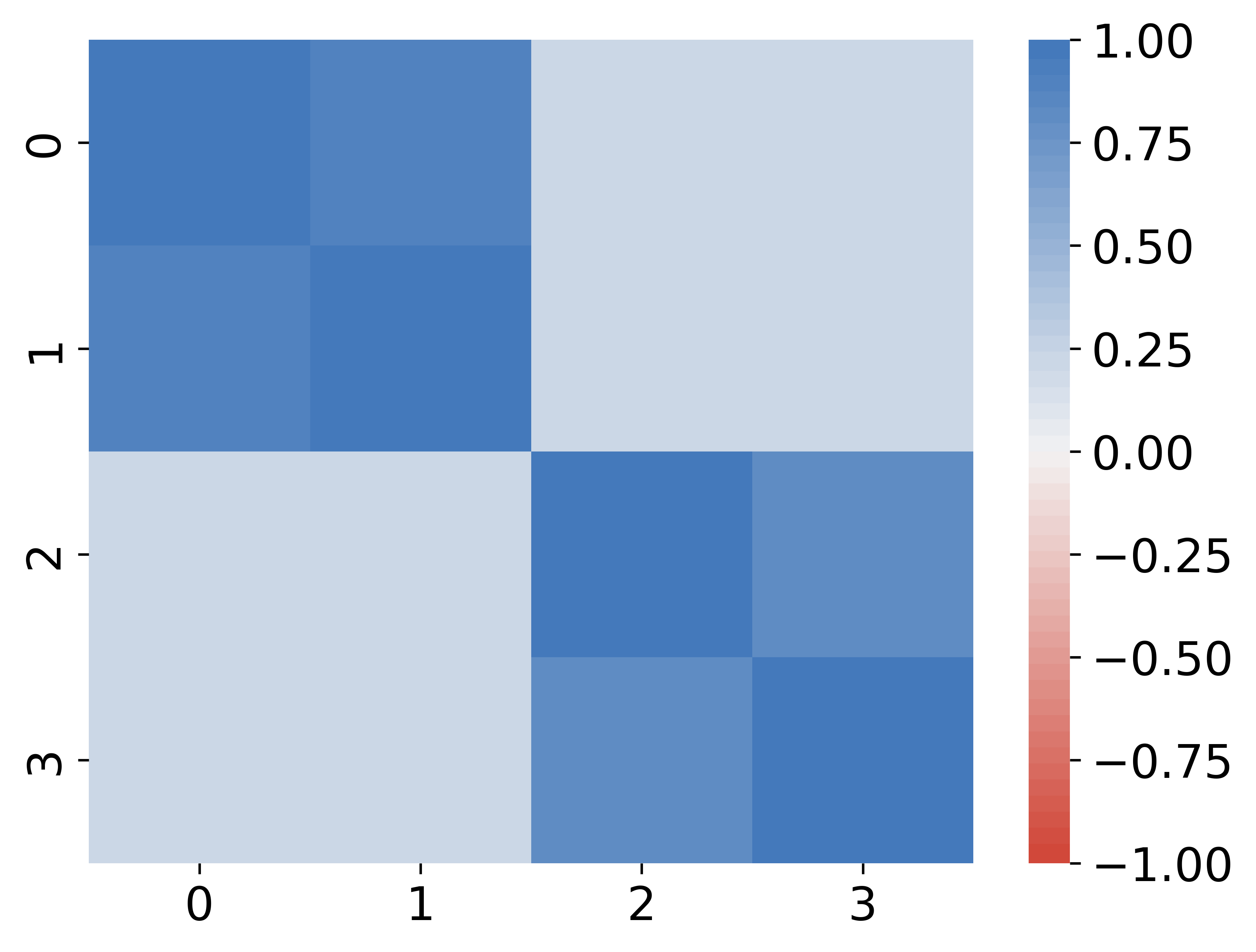}
         \caption{}
     \end{subfigure}
        \caption{The effect of applying  the hierarchical clustering estimator under the ALCA approach to a $4\times4$ matrix with a  hierarchical nested structure. (a) represent an empirical matrix $\mathbf{E}$, (b) the associated dendrogram, and (c) the filtered correlation matrix $\mathbf{\Xi}(\mathbf{E})$.}
        \label{Fig1}
\end{figure}

The hierarchical clustering estimator was proposed in~\cite{tumminello2007kullback}. 
This estimator is based on the hierarchical clustering methods, which require a distance or dissimilarity matrix as an input.
Then, we must transform the correlation matrix $\mathbf{E}$ into a dissimilarity matrix. 
Here, we choose the transformation $D_{ij}=1-E_{ij}$, which satisfies the axioms of a distance measure.
The clusters can generally be created through \emph{divisive} or \emph{agglomerative} methods.
The proposed estimator considers the agglomerative strategy, which consists of four steps.
The first step is to set each of the $p$ variables in a single cluster.
Next, in the second step, we search in $D$ for the nearest~(most similar) pair of variables~(clusters), say $a,b$, and denote this distance by $d_{ab}$. 
In the third step, the clusters $a$ and $b$ are merged, denoted as $(ab)$, and the entries of $D$ are updated by removing the rows and columns corresponding to the variables $a$ and $b$.  
Hence  the row and column regarding the new cluster (ab) distances to each of the remaining clusters are added to $D$.
The four-step consists of repeating  steps 1, 2, and 3 until a single cluster is obtained, where the levels at which two clusters join together can be represented through a dendrogram.
To quantify the nearest (most similar) clusters, we use the average linkage given by the agglomerative criteria~\cite{johnson2002applied}
\begin{equation}
d_{ab}  = \frac{\sum_i \sum_j D_{ij}}{N_{a}N_b}  
\end{equation}
where $D_{ij}$ consider the distance between the objects $i$ and $j$ on the clusters $a$ and $b$, respectively, and $N_{a}$,$N_b$ represent their number of items.
This particular procedure, known as the  Average Linkage Clustering Analysis~(ALCA), enables to compute the \emph{Cophenetic distance} $\rho$ on the associated \emph{dendrogram}~\cite{anderberg2014cluster}, which is the distance between clusters at each hierarchical level.
Finally, it is built up a dissimilarity matrix as a function of $\rho$: $\mathbf{D}(\rho)$; and the filtered correlation matrix is obtained by
$\mathbf{\Xi}(E)_{ij} = 1-D_{ij}(\rho)$. 
Figure~\ref{Fig1} schematically shows the mechanism of applying the hierarchical clustering estimator under the ALCA approach to a  $4\times4$ matrix with a hierarchical nested structure. The example shows that the finer structures are filtered out using this procedure, and only the strongest correlation blocks are preserved.

\subsection{Two-step estimators}  
We introduce the two-step estimators, which consist in applying as a first step a RIE estimator to deal with the statistical uncertainty due to high dimensionality. Once this type of uncertainty is eliminated, a hierarchical clustering-based estimator is applied as a second step to highlight the hierarchically nested  block structure. 
We expect that a two-step estimator present very good performance for several loss functions because (i) a first application of a rotationally invariant estimator reduces the error of estimation of largest eigenvalues and (ii) the second application of an appropriate filtering by hierarchical clustering can reduce the inconsistency in top eigenvectors that is unavoidably associated with rotationally invariant estimators.    
In particular, we consider the following combination of estimators because they present very good performance
\begin{itemize}
    \item  2-step (I): $\mathbf{\Xi}^{ALCA}(\mathbf{\Xi}^{mwcv}(\mathbf{E}))$ 
    \item 2-step (II): $\mathbf{\Xi}^{ALCA}(\mathbf{\Xi}^{BJ}(\mathbf{E}))$ 
    \item 2-step (III): $\mathbf{\Xi}^{ALCA}(\mathbf{\Xi}^{LP}(\mathbf{E}))$ 

\end{itemize}

\section{Loss Functions}
\label{LossFunctions}

We use six different loss functions to compare the effect of the different estimators on the correlation matrices. The first of these is the Kullback-Leibler divergence.

Let $\mathbf{A},\mathbf{B}$ be two square matrices of dimension $p\times p$, the KL divergence of Gaussian processes is given by ~\cite{tumminello2007kullback}
    \begin{equation}
        K(\mathbf{A},\mathbf{B}) = \frac{1}{2} \left[ \log [det (\mathbf{B} \mathbf{A}^{-1} )]+Tr(\mathbf{B}^{-1}\mathbf{A})-p\right].
    \end{equation}
We note that, under the assumption of Gaussianity, $K(\mathbf{A},\mathbf{B})$ is equivalent up to a factor to the commonly known inverse Stein's loss function~\cite{ledoit2018optimal}.

The second metric is the inverse KL divergence or Stein's loss. This metric has the same expression as the KL divergence given above but applied on the inverse matrices: $K(\mathbf{A}^{-1},\mathbf{B}^{-1})$. It is important to mention that Stein's loss function $\mathcal{L}^{Stein}$ is related to the inverse KL divergence by a scaling factor
\begin{eqnarray}
&& \mathcal{L}^{Stein}(\mathbf{A},\mathbf{B})  = \frac{1}{p} Tr \left( \mathbf{A}^{-1}\mathbf{B} \right) \\
&& - \frac{1}{p} \log [det (\mathbf{A}^{-1} \mathbf{B})]-1 =
\frac{2}{p} K(\mathbf{A}^{-1}, \mathbf{B}^{-1})
\end{eqnarray}
This work considers the scaled version to prevent the loss function from going to infinity with the matrix dimension, and both metrics (Stein's loss and inverse KL) are assumed to be indistinguishable. In~\cite{tumminello2007kullback} has been shown that the expected value of the KL divergence does not depend on the specific model under the Gaussian assumption. Consider two independent sample covariance matrices $\mathbf{E}_1, \mathbf{E}_2$ coming from the parent population $\mathbf{C}$, the next scaled expectations are valid under Gaussian assumptions
\begin{equation}
\begin{split}
    &\mathbb{E}[K(\mathbf{E}_1, \mathbf{E}_2)] = \frac{p+1}{n-p-1}\\
    &\mathbb{E}[K(\mathbf{C}, \mathbf{E})] = \\
    &\frac{1}{p}\left[p\log\left(\frac{2}{n}\right) 
    + \sum_{t=n-p+1}^n \left(\frac{\Gamma'(t/2)}{\Gamma(t/2)}\right)+\frac{p(p+1)}{n-p-1}\right]
    \label{scaled_expectations}
\end{split}
\end{equation}
where $\Gamma(x)$ is the usual gamma function and $\Gamma'(x)$ is the derivative of $\Gamma(x)$. We have scaled the metric by $\frac{2}{p}$ to have an exact equivalence between Stein's loss and the KL divergence, yet the original result does not consider this factor. 

Moreover, the Frobenius norm is given by
    \begin{equation}
        F(\mathbf{A},\mathbf{B}) = \frac{1}{p}Tr[(\mathbf{A}-\mathbf{B})(\mathbf{A}-\mathbf{B})'],
    \end{equation}
the corresponding inverse Frobenius is  $F(\mathbf{A}^{-1},\mathbf{B}^{-1})$.
The Frobenius and the inverse KL divergence are designed to deal with the covariance matrix, while the inverse Frobenius and the KL divergence to the inverse covariance matrix, also known as the precision matrix.

An interesting metric in the framework of the classical portfolio theory is the minimum-variance loss function~\cite{engle2019large}:
    \begin{equation}
        MV(\mathbf{A},\mathbf{B}) = \frac{Tr(\mathbf{B}^{-1} \mathbf{A} \mathbf{B}^{-1})/p}{[Tr(\mathbf{B}^{-1})/p]^2} - \frac{1}{Tr(\mathbf{A}^{-1})/p}
    \end{equation}
 
One last metric is the symmetrized Stein's loss, a combination of Stein's loss and the inverse of Stein's loss
    \begin{equation}
        SS(\mathbf{A},\mathbf{B}) = \frac{1}{p}Tr(\mathbf{B} \mathbf{A}^{-1} + \mathbf{B}^{-1} \mathbf{A}) - 2.
    \end{equation}
This metric pays equal attention to the problem of estimating the covariance and the precision matrix.

\section{Model}
We consider a multiplicative noise model with the following structure
\begin{eqnarray}
    \mathbf{Y} = \sqrt{\mathbf{C}}\mathbf{X} \sqrt{\mathbf{A}}\\
    \mathbf{E} = \frac{1}{n} \sqrt{\mathbf{C}} \mathbf{X}\mathbf{A}\mathbf{X}' \sqrt{\mathbf{C}}
    \label{noise_model}
\end{eqnarray}
where $\mathbf{Y}$ is the $p\times n$ data matrix, $\mathbf{C}$ is the $p\times p$ population cross-correlation matrix, $\mathbf{A}$ es de $n\times n$ autocorrelation matrix, and $X_{ij}\sim \mathcal{N}(0,1)$, that is, each element $X_{ij}$ follows a standard Gaussian distribution. 
The correlation model $\mathbf{C}$ is first constructed as follows
\begin{equation}
    \mathbf{L}_{kl} = 
    \begin{cases}
      \gamma_{kl}, & \text{if}\ k=k(l),\dots, k(l+p_l) \\
      0, & \text{otherwise},
    \end{cases}
\end{equation}
where $\mathbf{L}$ is the loading matrix of dimension $p\times b$, $k=1,\dots, p$, $l=1,\dots, b$, $p_l$ the size of each block $l$, being $b$ the number of blocks~($b\leq p$), and $\{k(l),k(l+p_l)\}$ are the initial and last value of the given block $l$. Once defined $\mathbf{L}$, the population correlation matrix $\mathbf{C}$ is obtained simply by the expressions
\begin{eqnarray}
    \mathbf{Q} &=& \mathbf{L}\mathbf{L}',\\
    C_{ij} &=&  \delta_{ij} + Q_{ij}(1-\delta_{ij}), 
\end{eqnarray}
where $\delta_{ij}$ denotes the Kronecker delta. 
We have considered a block diagonal and hierarchical nested block matrix structure to model $\mathbf{C}$. 
The first model comprises 12 independent diagonal blocks, while the second model is constructed with 12 overlapped diagonal blocks. 
In particular, we consider a homogeneous specification of the loading factors $\gamma_{kl}=\gamma=0.3$. 
In both models, the block sizes $p_l$ are heterogeneous as well as the initial and last values $\{k(l),k(l+p_l)\}$.

We analyze three different cases. The first case considers the block diagonal model with autocorrelation matrix $\mathbf{A} = \mathbf{I}$. The second is the hierarchical nested model with autocorrelation matrix $\mathbf{A} = \mathbf{I}$. And the third is the same hierarchical nested model but with autocorrelation elements of the form $\mathbf{A}_{ij}= e^{-\frac{|i-j|}{\tau}}$, where we have fixed $\tau=3$. In other words, the first two cases represent time series without memory, while in the third case, the memory decays exponentially as a function of the separation between the observations $i,j$. 

\section{Results}

\begin{figure}[!ht]
     \begin{subfigure}[b]{0.45\textwidth}
         \centering
        \includegraphics[scale=.6]{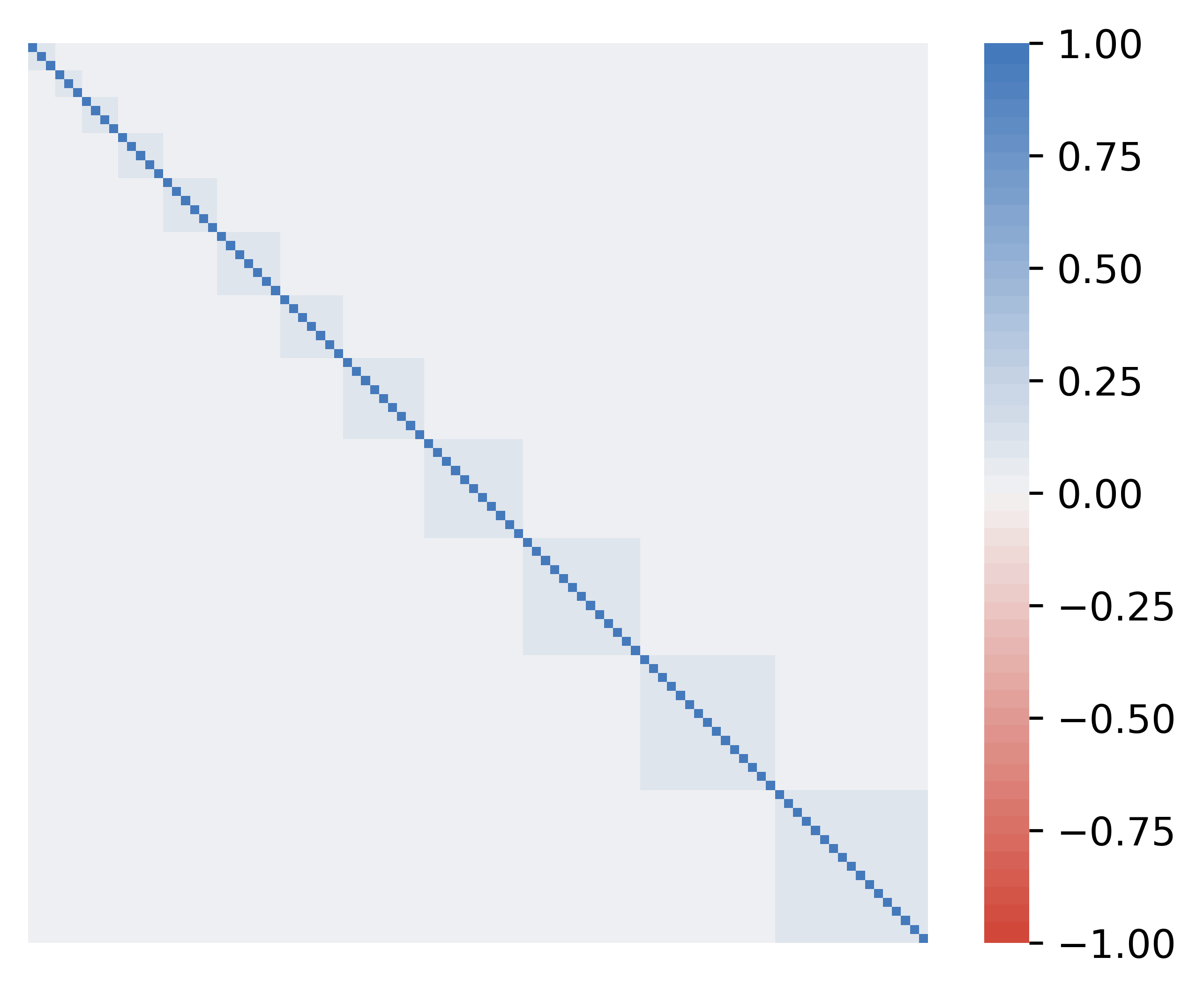}
         \caption{}
     \end{subfigure}
    \begin{subfigure}[b]{0.45\textwidth}
         \centering
        \includegraphics[scale=.6]{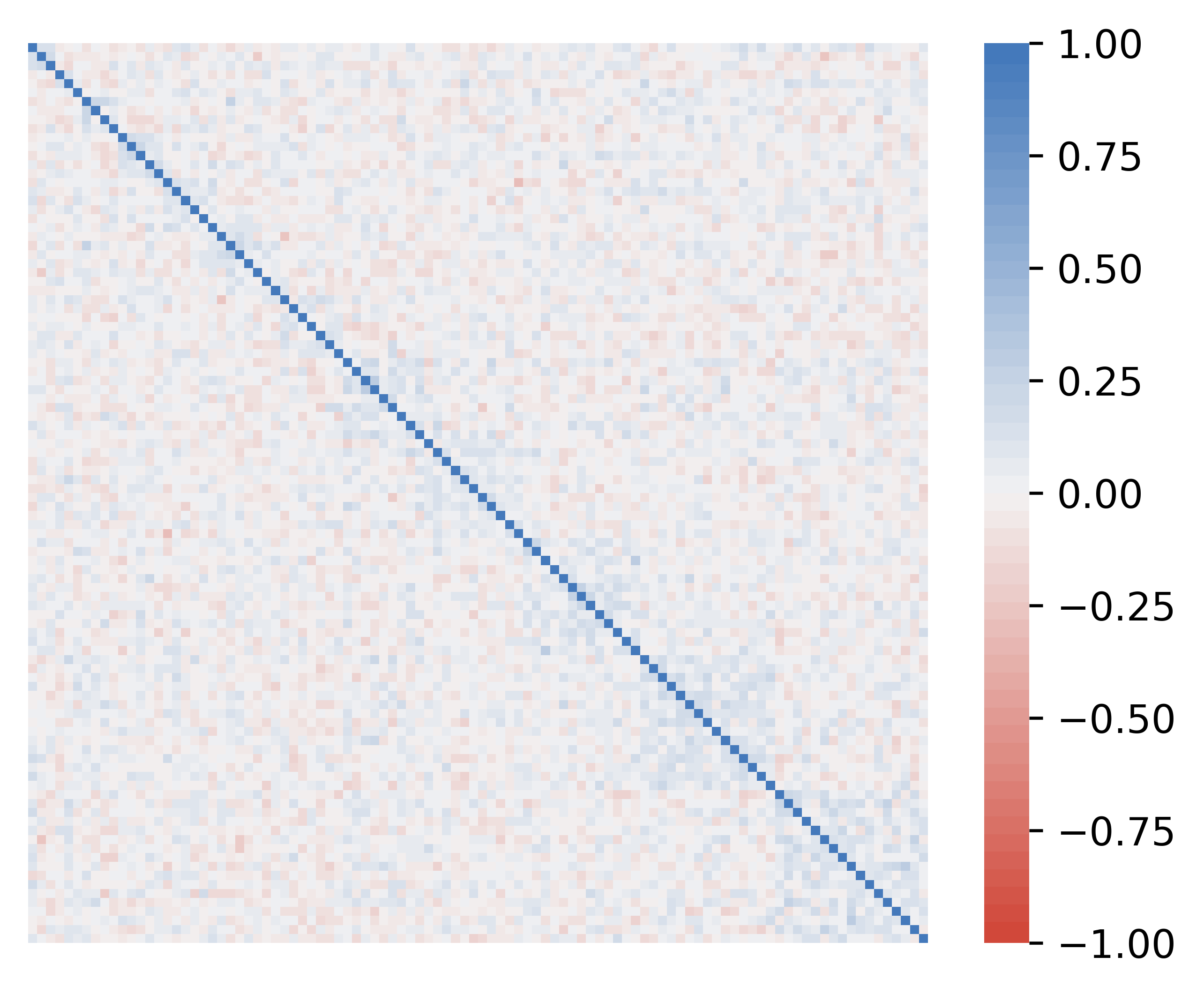}
         \caption{}
     \end{subfigure}
     \caption{Block diagonal model.
        (a) Population correlation matrix. (b) A single realization of such processes with autocorrelation matrix $\mathbf{A} = \mathbf{I}$ and dimensions $p=100,n=200$.}
        \label{Fig2}
\end{figure}

\begin{figure}[!ht]
     \begin{subfigure}[b]{0.45\textwidth}
         \centering
        \includegraphics[scale=.6]{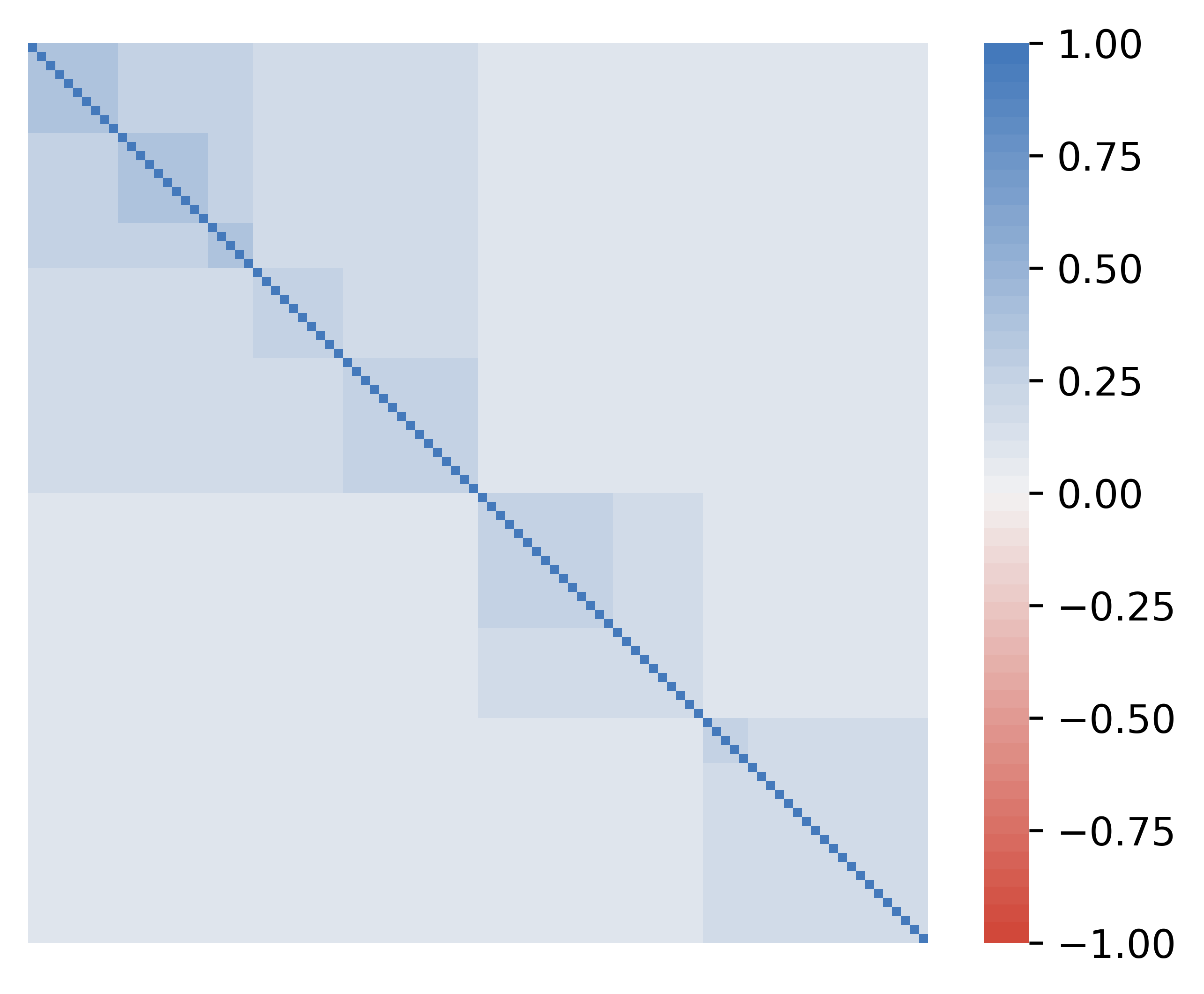}
         \caption{}
     \end{subfigure}
    \begin{subfigure}[b]{0.45\textwidth}
         \centering
        \includegraphics[scale=.6]{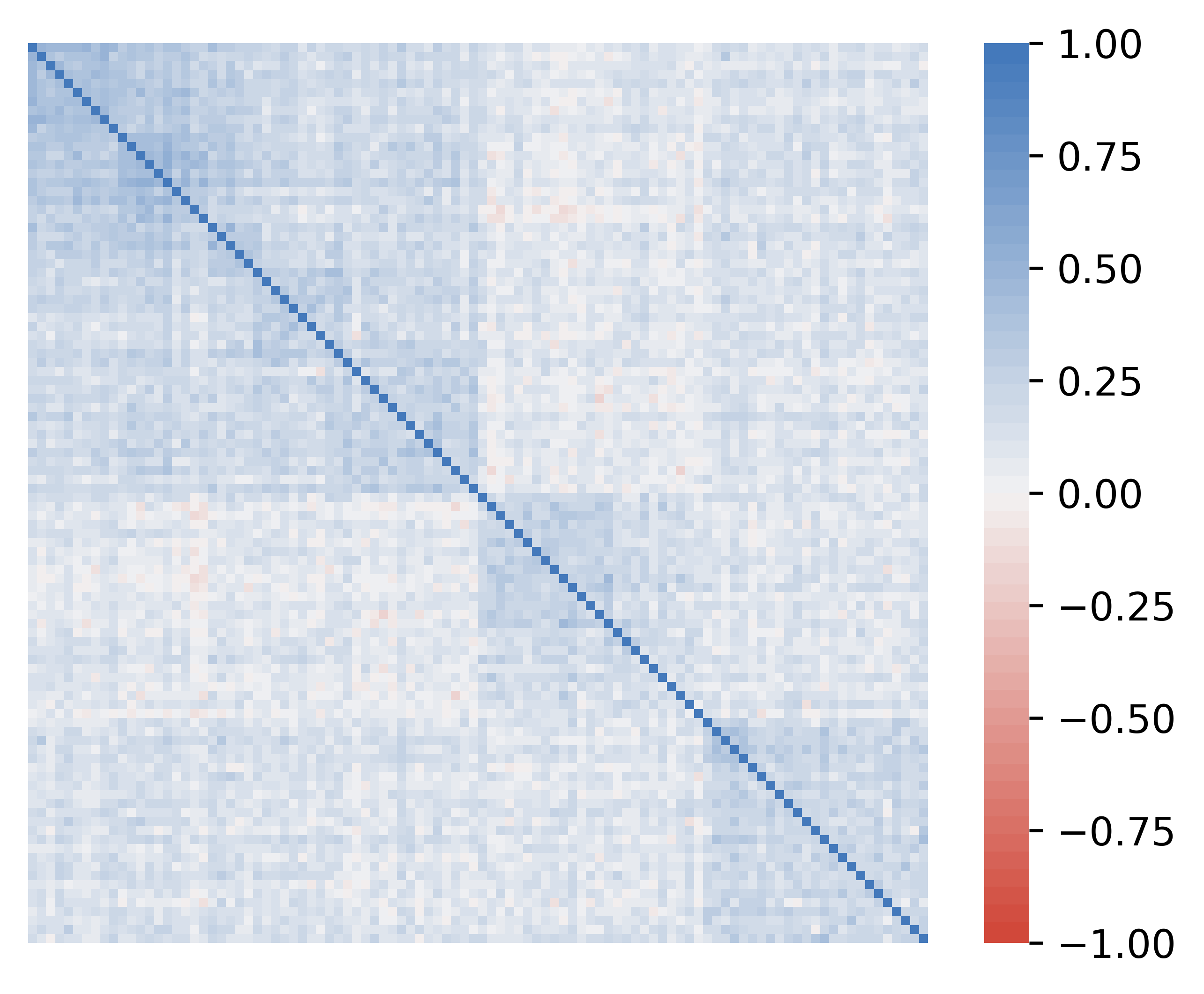}
         \caption{}
     \end{subfigure}
     \begin{subfigure}[b]{0.45\textwidth}
         \centering
        \includegraphics[scale=.6]{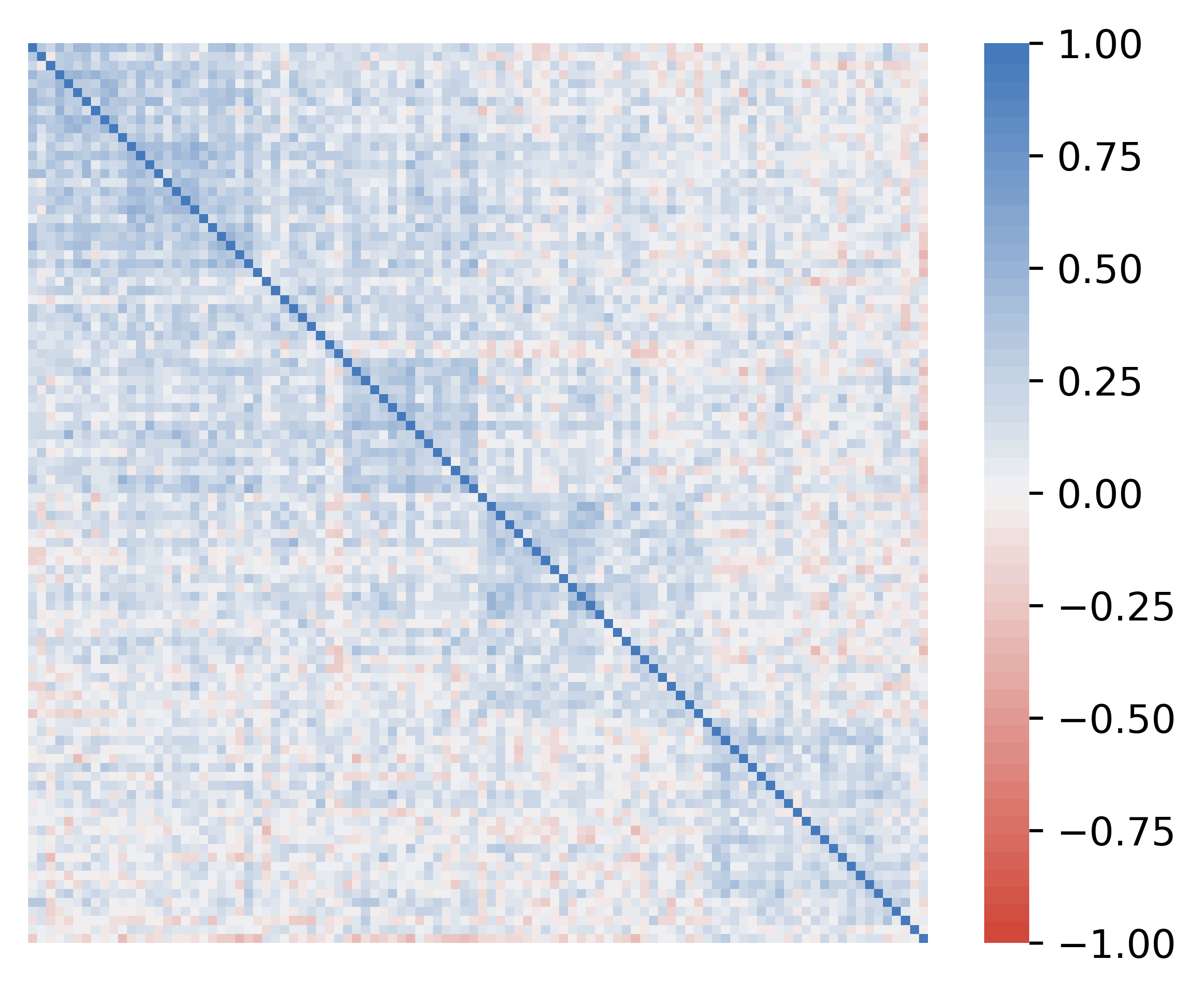}
         \caption{}
     \end{subfigure}
        \caption{Hierarchical nested model.
        (a) Population correlation matrix. (b) A single realization of such processes with autocorrelation matrix $\mathbf{A} = \mathbf{I}$.  (c) A single realization of such processes with autocorrelation elements $\mathbf{A}_{ij}= e^{-|i-j|/3}$.  The dimensions of the samples are $p=100,n=200$.}
        \label{Fig3}
\end{figure}

We generate $m$ realizations of multivariate time series $\mathbf{Y}$ for each study case. Each sample matrix $\mathbf{E}$ is computed and filtered using estimators described in Section~\ref{estimators}. Subsequently, the estimator's performance is measured through the six loss functions described in section~\ref{LossFunctions}. 
Figure \ref{Fig2} shows a graphical representation of the block diagonal model accompanied by a single realization of the process with $\mathbf{A} = \mathbf{I}$ (case 1). Likewise, figure \ref{Fig3} shows a graphical representation of the hierarchical nested model accompanied by a single realization of the process with $\mathbf{A} = \mathbf{I}$ (case 2) and with autocorrelation elements of the form $\mathbf{A}_{ij}= e^{-|i-j|/3}$ (case 3). The realizations are made for dimensions $p=100$ and $n=200$.
It can be seen that sample matrices show statistical uncertainty because the number of observations and the number of variables is finite. The noise occurs naturally when we study cross-correlations of a large number of variables with a limited number of records. This condition is quite common in many research fields. For example, practitioners in finance prefer a high dimensional setting, i.e., $p \sim n$, to avoid non-stationary effects or structural changes in return time series of assets traded in financial markets.

\begin{figure*}[!ht]
    \centering
    \begin{subfigure}[b]{0.45\textwidth}
        \includegraphics[scale=.3]{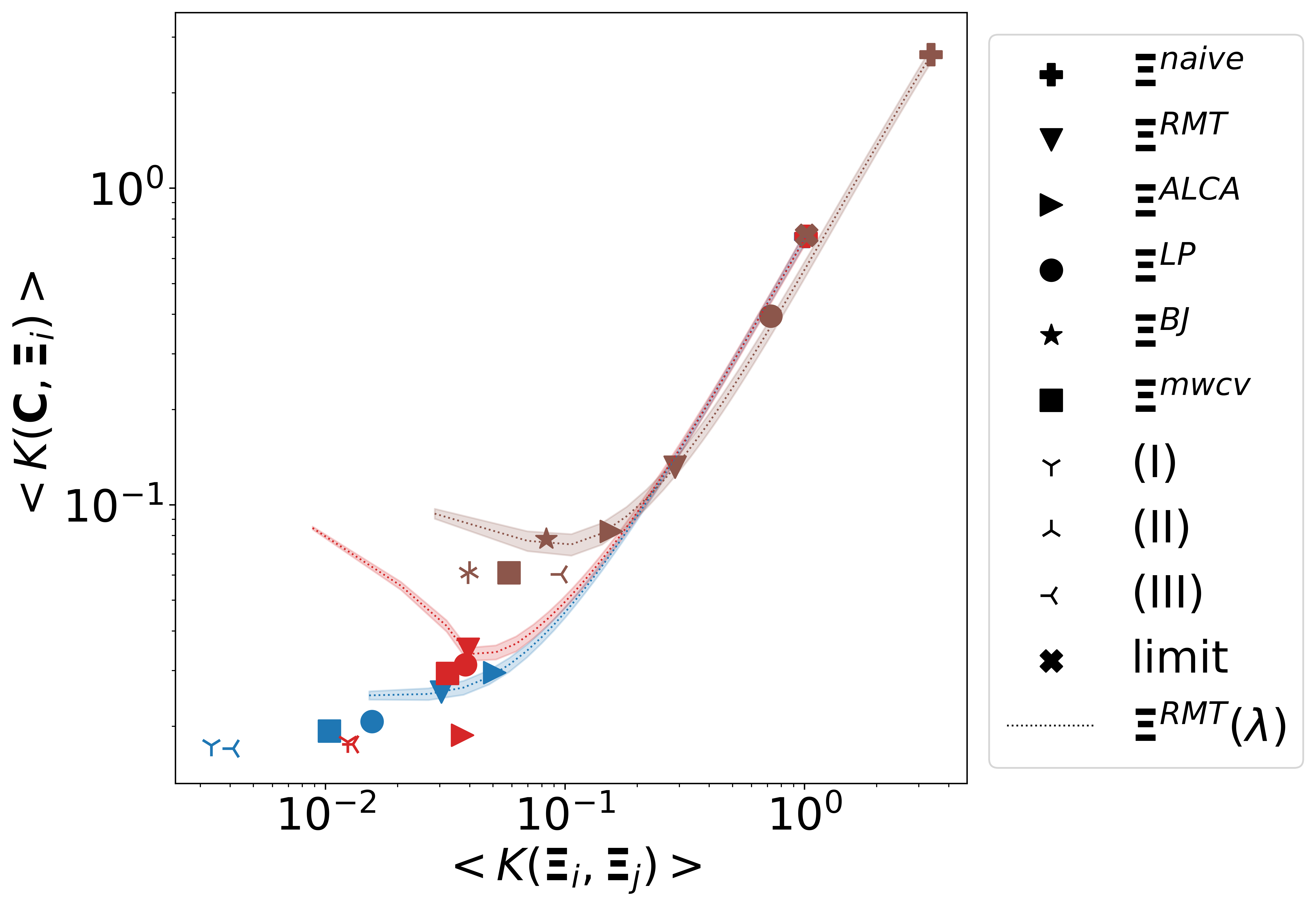}
    \caption{}
    \end{subfigure}
    \begin{subfigure}[b]{0.45\textwidth}
        \includegraphics[scale=.3]{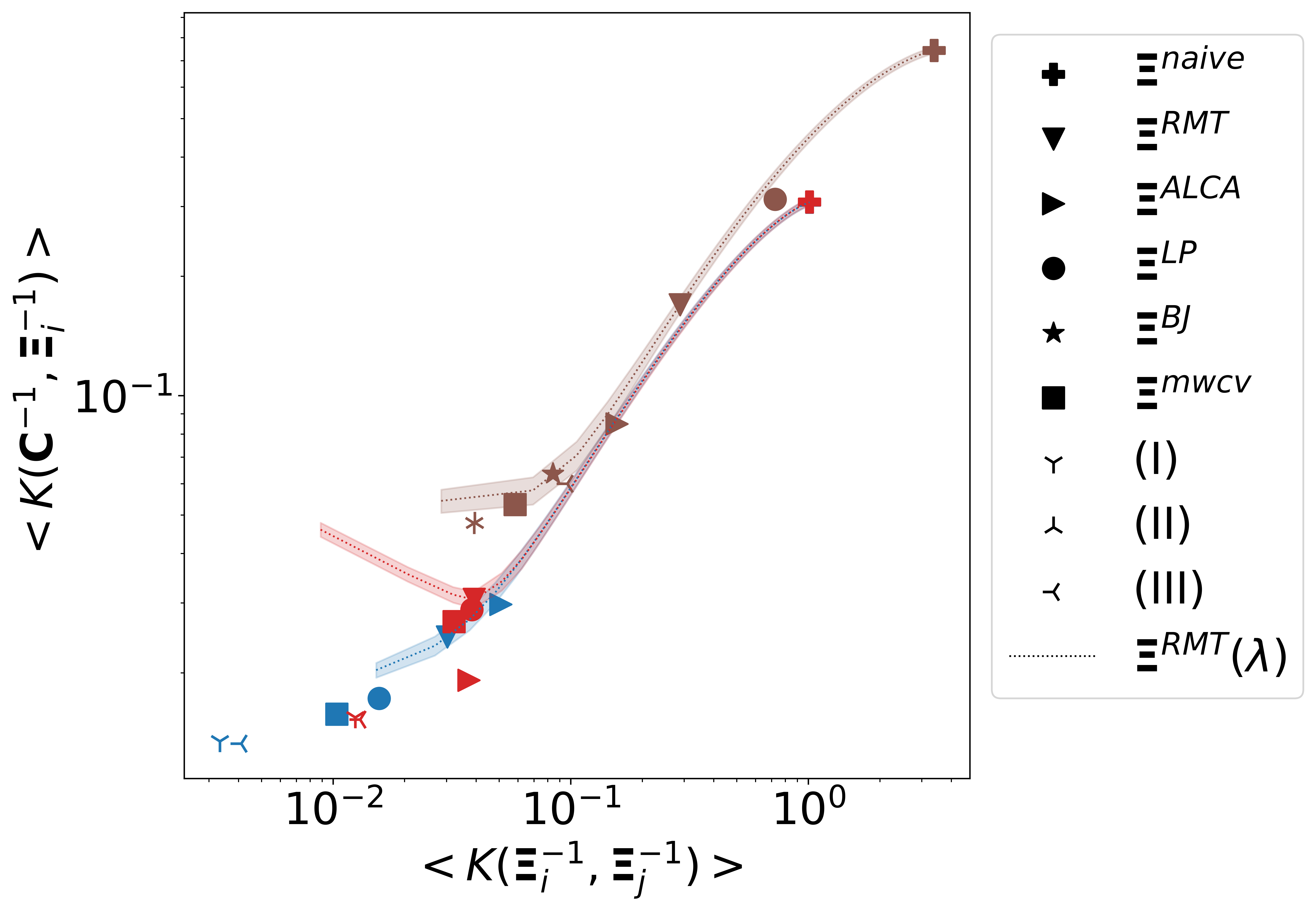}
    \caption{}
    \end{subfigure}\\
        \begin{subfigure}[b]{0.45\textwidth}
        \includegraphics[scale=.3]{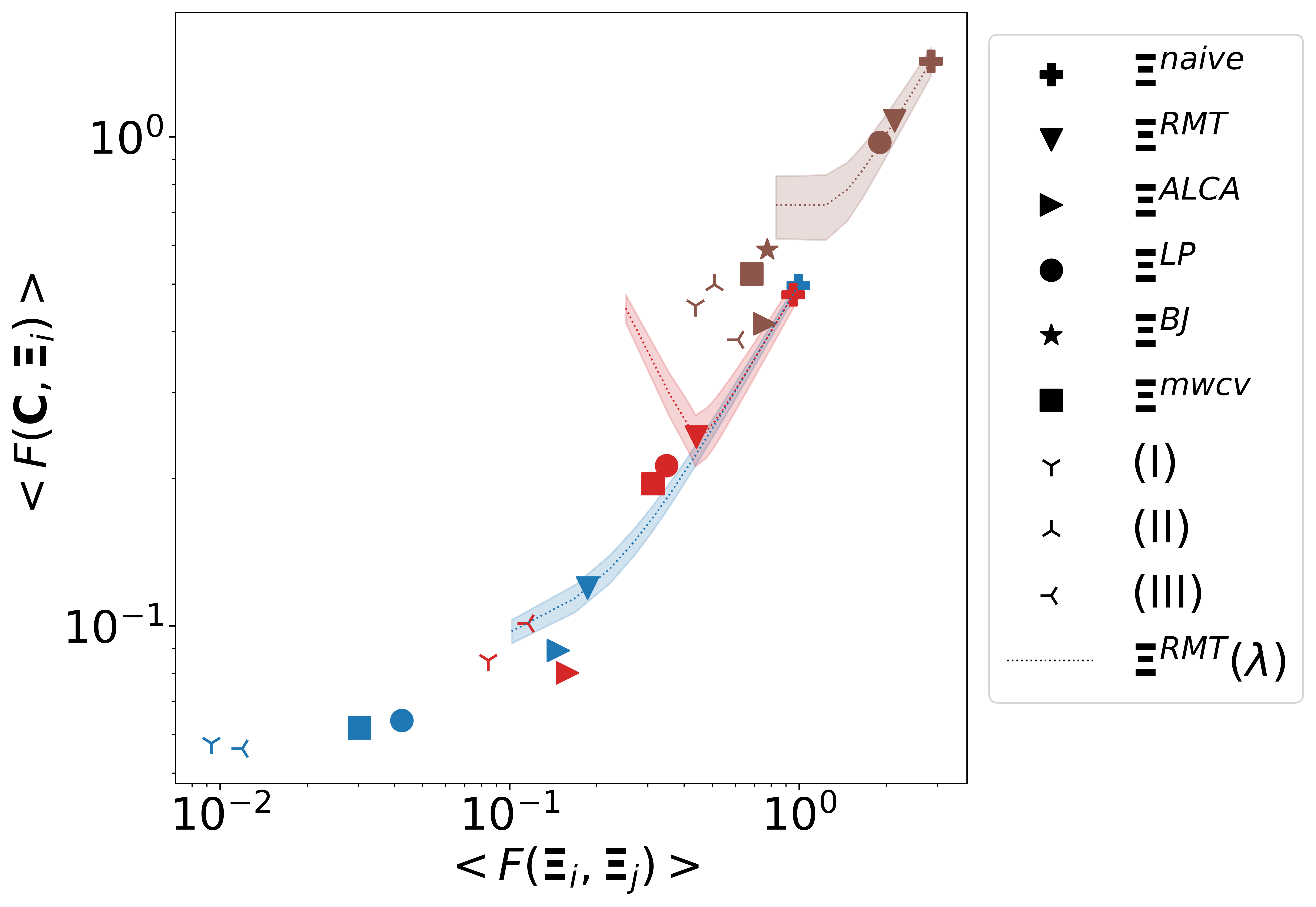}
    \caption{}
    \end{subfigure}
    \begin{subfigure}[b]{0.45\textwidth}
        \includegraphics[scale=.3]{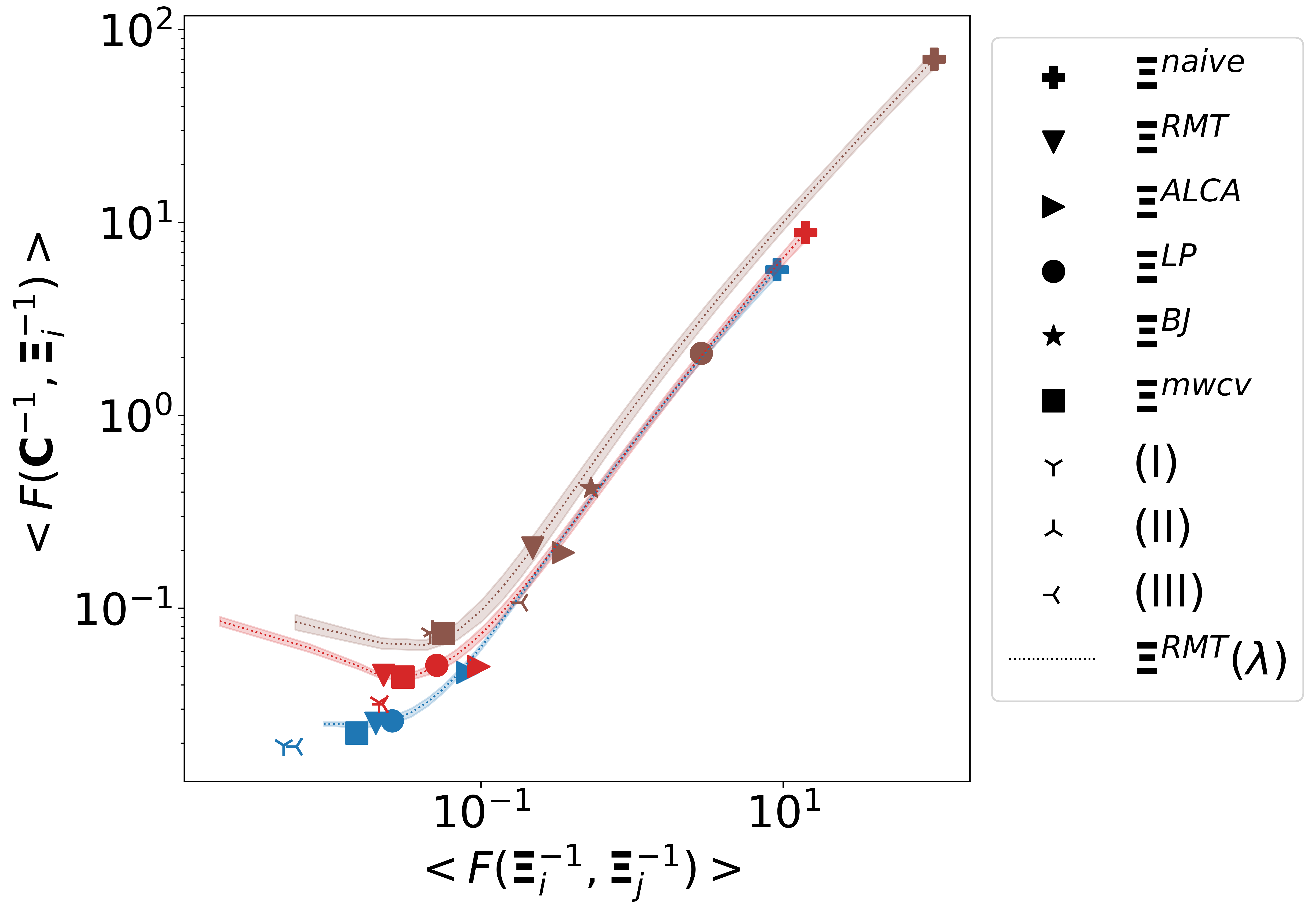}
    \caption{}
    \end{subfigure}\\
    \begin{subfigure}[b]{0.45\textwidth}
        \includegraphics[scale=.3]{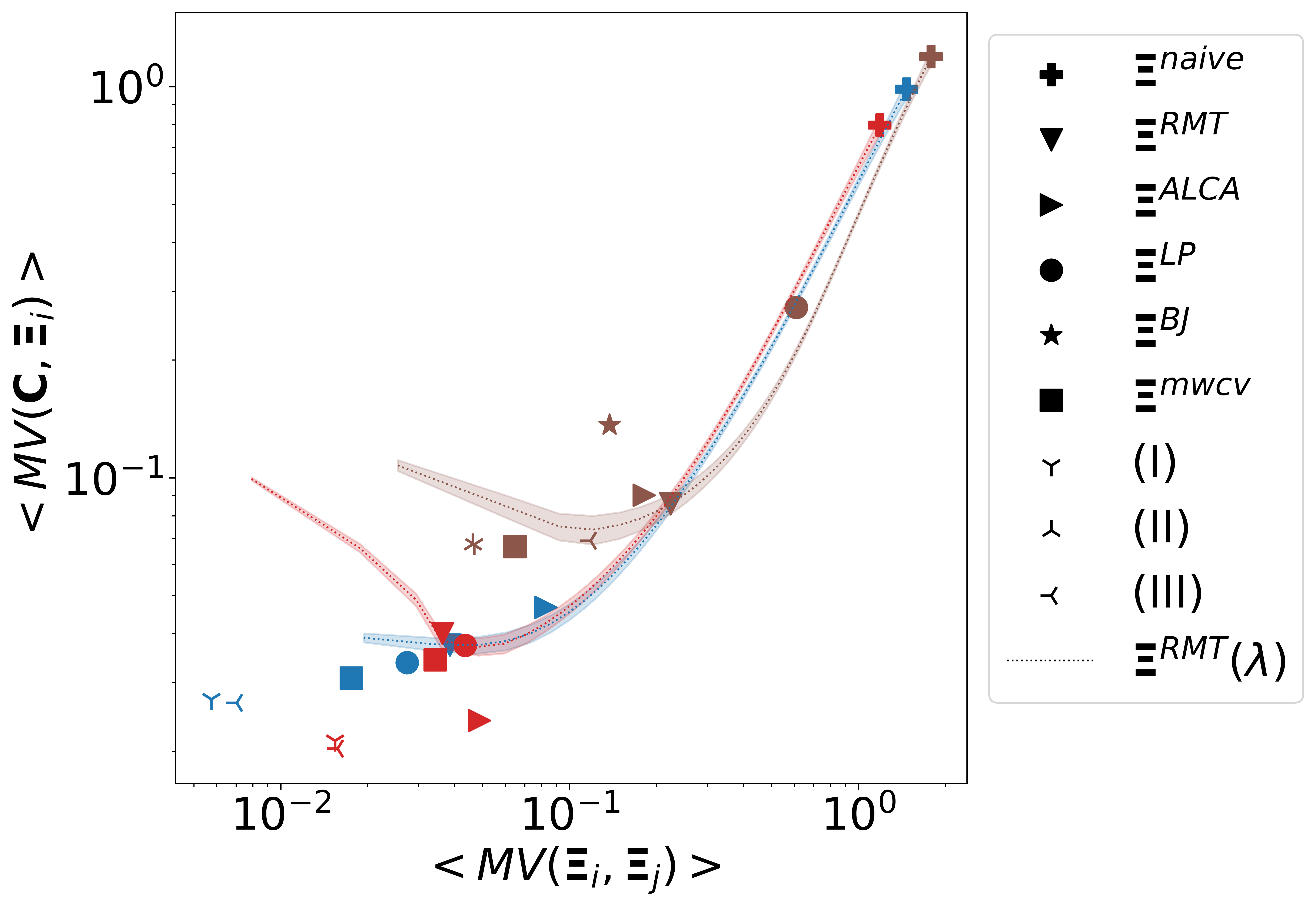}
    \caption{}
    \end{subfigure}
    \begin{subfigure}[b]{0.45\textwidth}
        \includegraphics[scale=.3]{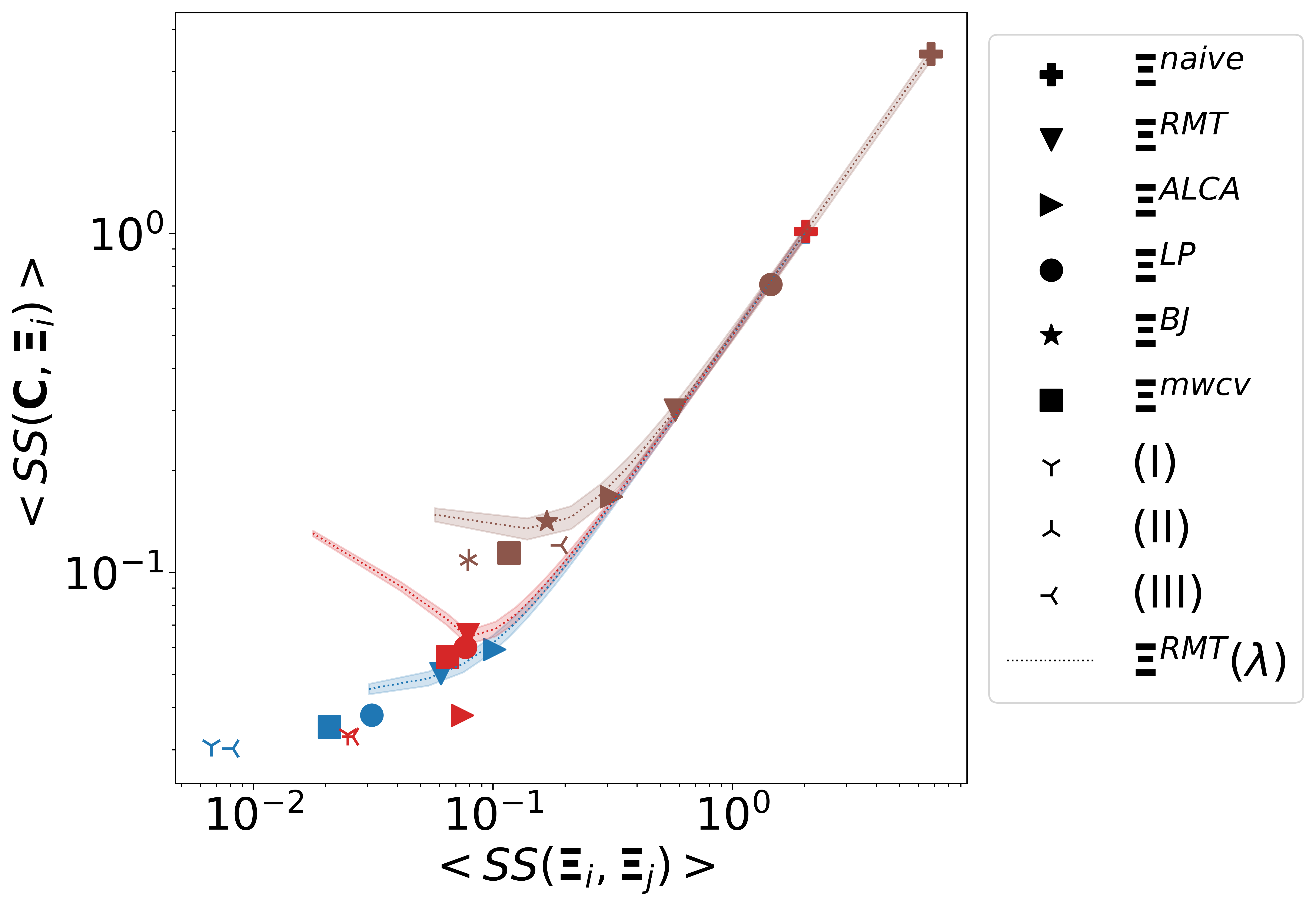}
    \caption{}
    \end{subfigure}
    \caption{Average loss functions over $m=1000$ realizations of the multiplicative noise model~(see Eq.~\ref{noise_model} ) for dimensions $p=100, n=200$. The block diagonal model without memory (case 1) is represented by blue, the hierarchical nested model without memory (case 2) is represented by red, and the hierarchical nested model with memory (case 3) is represented by brown. (a) $\langle K(\mathbf{C},\mathbf{\Xi}_i)\rangle$ vs. $\langle K(\mathbf{\Xi}_i,\mathbf{\Xi}_j)\rangle$, where the cross-marker represent the theoretical limits given by eq.~\ref{scaled_expectations}. (b) $\langle K(\mathbf{C}^{-1},\mathbf{\Xi}_i^{-1})\rangle$ vs. $\langle K(\mathbf{\Xi}_i^{-1},\mathbf{\Xi}_j^{-1})\rangle$. (c) $\langle F(\mathbf{C},\mathbf{\Xi}_i)\rangle$ vs. $\langle F(\mathbf{\Xi}_i,\mathbf{\Xi}_j)\rangle$. (d) $\langle F(\mathbf{C}^{-1},\mathbf{\Xi}_i^{-1})\rangle$ vs. $\langle F(\mathbf{\Xi}_i^{-1},\mathbf{\Xi}_j^{-1})\rangle$. (e) $\langle MV(\mathbf{C},\mathbf{\Xi}_i)\rangle$ vs. $\langle MV(\mathbf{\Xi}_i,\mathbf{\Xi}_j)\rangle$. (f) $\langle SS(\mathbf{C},\mathbf{\Xi}_i)\rangle$ vs. $\langle SS(\mathbf{\Xi}_i,\mathbf{\Xi}_j)\rangle$. Both axes are on a logarithmic scale.}
    \label{Fig4}
\end{figure*}

Figure~\ref{Fig4} shows the behavior of the average loss functions over  $m=1000$ realizations and dimensions $p=100, n=200$ for case 1~(blue), case 2 (green), and case 3 (brown). Panels from (a) to (f) show  $\langle\mathcal{L}(\mathbf{C},\mathbf{\Xi}_i)\rangle$ vs. $\langle \mathcal{L}(\mathbf{\Xi}_i,\mathbf{\Xi}_j)\rangle$. We denote by $\mathcal{L}$ each of the loss functions~(KL divergence, Frobenius, etc.), $\langle \cdot\rangle$ represents the average, and $\mathbf{\Xi}$ represents the filtered correlation matrix under each of the filtering strategies described in Section~\ref{estimators}.
The $\mathbf{\mathbf{\Xi}}^{mwcv}$ estimator is set with $T_{total}=10T$ and $T_{out}=T=n$. 
Moreover, the $\Xi^{BJ}$ estimator is set with $\tau=3$~(or equivalently $\eta=\coth(1/3)\approx 3.11$; see eqs. \ref{BurdaJarosz2022} and \ref{Zeta}). Under this setting, the $\Xi^{BJ}$ filter is expected to obtain optimal results for case 3, while the filter would be misspecified to deal with cases 1 and 2. Thus, we have omitted the results under the $\Xi^{BJ}$ and the related 2-step (II) estimator for the latter cases.
In these curves, we are comparing the value that minimizes the loss function (y-axis) given the stability level of the estimators (x-axis).
The dotted line represents the average RMT estimator when the number of eigenvalues $\lambda$ that are kept in the filtering procedure varies from $1$ to $p$, where the shadow band represents the standard deviation. The lower left corner corresponds to the case where we have kept only the signal associated with the largest eigenvalue. The upper right corner corresponds to the extreme case where we have kept all the signals or eigenvalues, so the estimator is identical to the empirical correlation matrix $\mathbf{E}$.  
Our simulations confirm that $\langle K(\mathbf{C},\mathbf{\Xi}_i^{naive})\rangle$ is in agreement with the theoretical limits of the KL divergence given by eqs.~\ref{scaled_expectations}~(represented by the cross-marker in panel (a)). 

The curves' behavior  of the block diagonal model~(blue color) is monotonically increasing almost for every value, except very near the origin. 
In contrast, the curves of the hierarchical nested model~(red and brown colors) are monotonically increasing only relatively far from the origin. 
Interestingly, the RMT filter roughly coincides with the numerical minimum of $\langle \mathcal{L}(\mathbf{C},\mathbf{\Xi}^{RMT}(\lambda))\rangle$~(dotted lines) for all the metrics $\mathcal{L}$ when no autocorrelations are considered (case 2).
Thus, the Marchenko-Pastur bound effectively gives us the number of optimal signals to preserve in the hierarchical nested model without autocorrelation but fails to recover the true number of signals if the model violates the i.i.d. assumption (case 3).
In general, we can see that the two-step estimators are the ones that obtain the optimal and most stable values within each case and for all the considered loss functions.

Tables~\ref{table1}, \ref{table2}, and \ref{table3} summarize the performance of estimators in terms of $\langle \mathcal{L}(\mathbf{C},\mathbf{\Xi}_i)\rangle$ for the three study~(see Appendix~\ref{apendice2})
The filter' stability  $\langle \mathcal{L}(\mathbf{\Xi}_i,\mathbf{\Xi}_j)\rangle$ of each estimator $\mathbf{\Xi}$ in relation to the loss function $\mathcal{L}$ are shown in Appendix~\ref{apendice2}~(see tables~\ref{table4},\ref{table5}, and \ref{table6}).
\begin{table*}[!ht]
\small
\caption{Block diagonal model without memory (case 1). Performance of estimators in terms of $\langle \mathcal{L}(\mathbf{C},\mathbf{\Xi}_i)\rangle$, where $\mathcal{L}$ denotes the loss function and $\langle \cdot\rangle$ represents the average over $m=1000$ realizations and considering dimensions $p=100, n=200$.}
\begin{tabular}{|l|r|r|r|r|r|r|}
\hline
 & \multicolumn{1}{l|}{$\langle K(\mathbf{C},\mathbf{\Xi}_i)\rangle$} & \multicolumn{1}{l|}{$\langle K(\mathbf{C}^{-1},\mathbf{\Xi}^{-1}_i)\rangle$} & \multicolumn{1}{l|}{$\langle F(\mathbf{C},\mathbf{\Xi}_i)\rangle$} & \multicolumn{1}{l|}{$\langle F(\mathbf{C}^{-1},\mathbf{\Xi}^{-1}_i)\rangle$} & \multicolumn{1}{l|}{$\langle MV(\mathbf{C},\mathbf{\Xi}_i)\rangle$} & \multicolumn{1}{l|}{$\langle SS(\mathbf{C},\mathbf{\Xi}_i)\rangle$} \\ \hline
$\mathbf{\Xi}^{naive}$ & 0.702978 & 0.307302 & 0.496612 & 5.682767 & 0.985036 & 1.010279 \\
$\mathbf{\Xi}^{RMT}$ & 0.025633 & 0.024646 & 0.119511 & 0.025346 & 0.037386 & 0.050279 \\
$\mathbf{\Xi}^{ALCA}$ & 0.029513 & 0.029777 & 0.089017 & 0.046473 & 0.046671 & 0.059290 \\
$\mathbf{\Xi}^{LP}$ & 0.020697 & 0.017244 & 0.064047 & 0.026045 & 0.033724 & 0.037942 \\
$\mathbf{\Xi}^{mwcv}$ & 0.019300 & 0.015717 & 0.061817 & 0.022541 & 0.030759 & 0.035017 \\
2-step (I) & 0.017404 & 0.013441 & 0.057510 & 0.019467 & 0.027199 & 0.030845 \\
2-step (III) & {\bf 0.017012} & {\bf 0.013252} & {\bf 0.056114} & {\bf 0.019154} & {\bf 0.026599} & {\bf 0.030264} \\

\hline
\end{tabular}
\label{table1}
\end{table*}
\begin{table*}[!ht]
\small
\caption{Hierarchical nested model without memory (case 2). Performance of estimators in terms of $\langle \mathcal{L}(\mathbf{C},\mathbf{\Xi}_i)\rangle$, where $\mathcal{L}$ denotes the loss function and $\langle \cdot\rangle$ represents the average over $m=1000$ realizations and considering dimensions $p=100, n=200$.}
\begin{tabular}{|l|r|r|r|r|r|r|}
\hline
 & \multicolumn{1}{l|}{$\langle K(\mathbf{C},\mathbf{\Xi}_i)\rangle$} & \multicolumn{1}{l|}{$\langle K(\mathbf{C}^{-1},\mathbf{\Xi}^{-1}_i)\rangle$} & \multicolumn{1}{l|}{$\langle F(\mathbf{C},\mathbf{\Xi}_i)\rangle$} & \multicolumn{1}{l|}{$\langle F(\mathbf{C}^{-1},\mathbf{\Xi}^{-1}_i)\rangle$} & \multicolumn{1}{l|}{$\langle MV(\mathbf{C},\mathbf{\Xi}_i)\rangle$} & \multicolumn{1}{l|}{$\langle SS(\mathbf{C},\mathbf{\Xi}_i)\rangle$} \\ \hline
$\mathbf{\Xi}^{naive}$ & 0.704715 & 0.308027 & 0.475366 & 8.871255 & 0.796652 & 1.012742 \\
$\mathbf{\Xi}^{RMT}$ & 0.035000 & 0.030668 & 0.243333 & 0.044913 & 0.040020 & 0.065668 \\
$\mathbf{\Xi}^{ALCA}$ & 0.018728 & 0.019151 & {\bf 0.080116} & 0.049903 & 0.023976 & 0.037879 \\
$\mathbf{\Xi}^{LP}$ & 0.031284 & 0.028860 & 0.212684 & 0.050627 & 0.037316 & 0.060144 \\
$\mathbf{\Xi}^{mwcv}$ & 0.029365 & 0.026917 & 0.195309 & 0.043923 & 0.034257 & 0.056282 \\
2-step (I) & 0.017844 & 0.015434 &  0.084999 & 0.032445 & 0.021296 & 0.033278 \\
2-step (III) & {\bf 0.017522} & {\bf 0.015251} & 0.101058 & {\bf 0.031783} & {\bf 0.020347} & {\bf 0.032772} \\

\hline
\end{tabular}
\label{table2}
\end{table*}
\begin{table*}[!ht]
\small
\caption{Hierarchical nested model with memory (case 3). Performance of estimators in terms of $\langle \mathcal{L}(\mathbf{C},\mathbf{\Xi}_i)\rangle$, where $\mathcal{L}$ denotes the loss function and $\langle \cdot\rangle$ represents the average over $m=1000$ realizations and considering dimensions $p=100, n=200$.}
\begin{tabular}{|l|r|r|r|r|r|r|}
\hline
 & \multicolumn{1}{l|}{$\langle K(\mathbf{C},\mathbf{\Xi}_i)\rangle$} & \multicolumn{1}{l|}{$\langle K(\mathbf{C}^{-1},\mathbf{\Xi}^{-1}_i)\rangle$} & \multicolumn{1}{l|}{$\langle F(\mathbf{C},\mathbf{\Xi}_i)\rangle$} & \multicolumn{1}{l|}{$\langle F(\mathbf{C}^{-1},\mathbf{\Xi}^{-1}_i)\rangle$} & \multicolumn{1}{l|}{$\langle MV(\mathbf{C},\mathbf{\Xi}_i)\rangle$} & \multicolumn{1}{l|}{$\langle SS(\mathbf{C},\mathbf{\Xi}_i)\rangle$} \\ \hline
$\mathbf{\Xi}^{naive}$ & 2.636596 & 0.741668 & 1.428074 & 70.128771 & 1.192546 & 3.378264 \\
$\mathbf{\Xi}^{RMT}$ & 0.131541 & 0.169630 & 1.078062 & 0.205236 & 0.085844 & 0.301171 \\
$\mathbf{\Xi}^{ALCA}$ & 0.082404 & 0.084835 & 0.414592 & 0.194150 & 0.090226 & 0.167239 \\
$\mathbf{\Xi}^{LP}$ & 0.394159 & 0.312629 & 0.974296 & 2.093030 & 0.272487 & 0.706789 \\
$\mathbf{\Xi}^{BJ}$ & 0.077930 & 0.063499 & 0.587735 & 0.419335 & 0.136441 & 0.141428 \\
$\mathbf{\Xi}^{mwcv}$ & 0.060966 & 0.053122 & 0.524506 & {\bf 0.073864} & {\bf 0.066712} & 0.114087 \\
2-step (I) & 0.060880 & {\bf 0.047672} & 0.451754 &  0.074127 &  0.067626 & {\bf 0.108552} \\
2-step (II) & 0.061383 & 0.047843 & 0.498786 & 0.076085 & 0.067689 & 0.109226 \\
2-step (III) & {\bf 0.060302} & 0.059961 & {\bf 0.384679} & 0.106745 & 0.069054 & 0.120263 \\

\hline
\end{tabular}
\label{table3}
\end{table*}

Table \ref{table1} shows the 2-step (III) estimator minimizes all the loss functions for case 1. In other words, the best strategy is to apply the $\mathbf{\Xi}^{LP}$ estimator followed by the  ALCA filter. The second best option is the 2-step (I) estimator, which implies applying the estimator $\mathbf{\Xi}^{mwcv}$ followed again by the ALCA filter. Notably, in third place, and very close to the minimum values of the two-step estimators mentioned above, are the results of simply applying the strategy $\mathbf{\Xi}^{mwcv}$.

The results for case 2 are similar for the two best performances.
The exception concerns the Frobenius metric, where now the ALCA filter beats the 2-step estimator (III) performance and moves it to third place, with the 2-step (I) estimator having the second-best performance against this metric. Actually, the ALCA estimator turn shifts $\mathbf{\Xi}^{mwcv}$ to obtain the third-best performance about the KL, inverse KL, MV, SS loss functions. Only with the inverse Frobeniuos metric does the estimator $\mathbf{\Xi}^{mwcv}$ obtain the third place.

For case 3, we have included the $\Xi^{BJ}$ filter, which systematically beats the $\Xi^{LP}$ filter as should be because it is calibrated with the same parameter $\tau=3$ of the generating process. Although the best performance is disputed between the $\Xi^{mwcv}$, 2-step(I), and 2-step(III) filters depending on the loss function. The top three also include the 2-step(II) and $\Xi^{ALCA}$ filters under some metrics.

\begin{figure}[!ht]
     \begin{subfigure}[b]{0.45\textwidth}
         \centering
        \includegraphics[scale=.3]{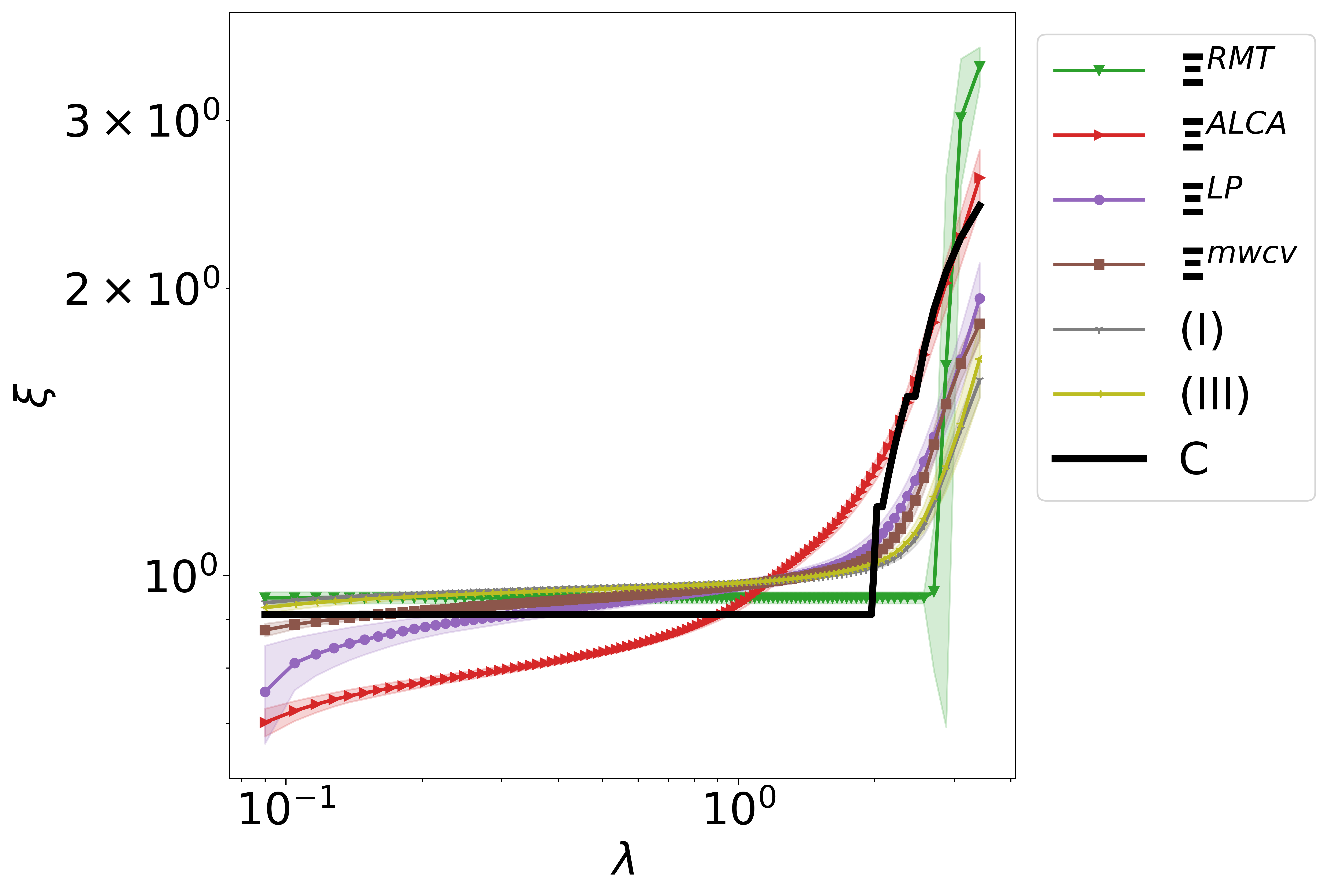}
         \caption{}
     \end{subfigure}
    \begin{subfigure}[b]{0.45\textwidth}
         \centering
        \includegraphics[scale=.3]{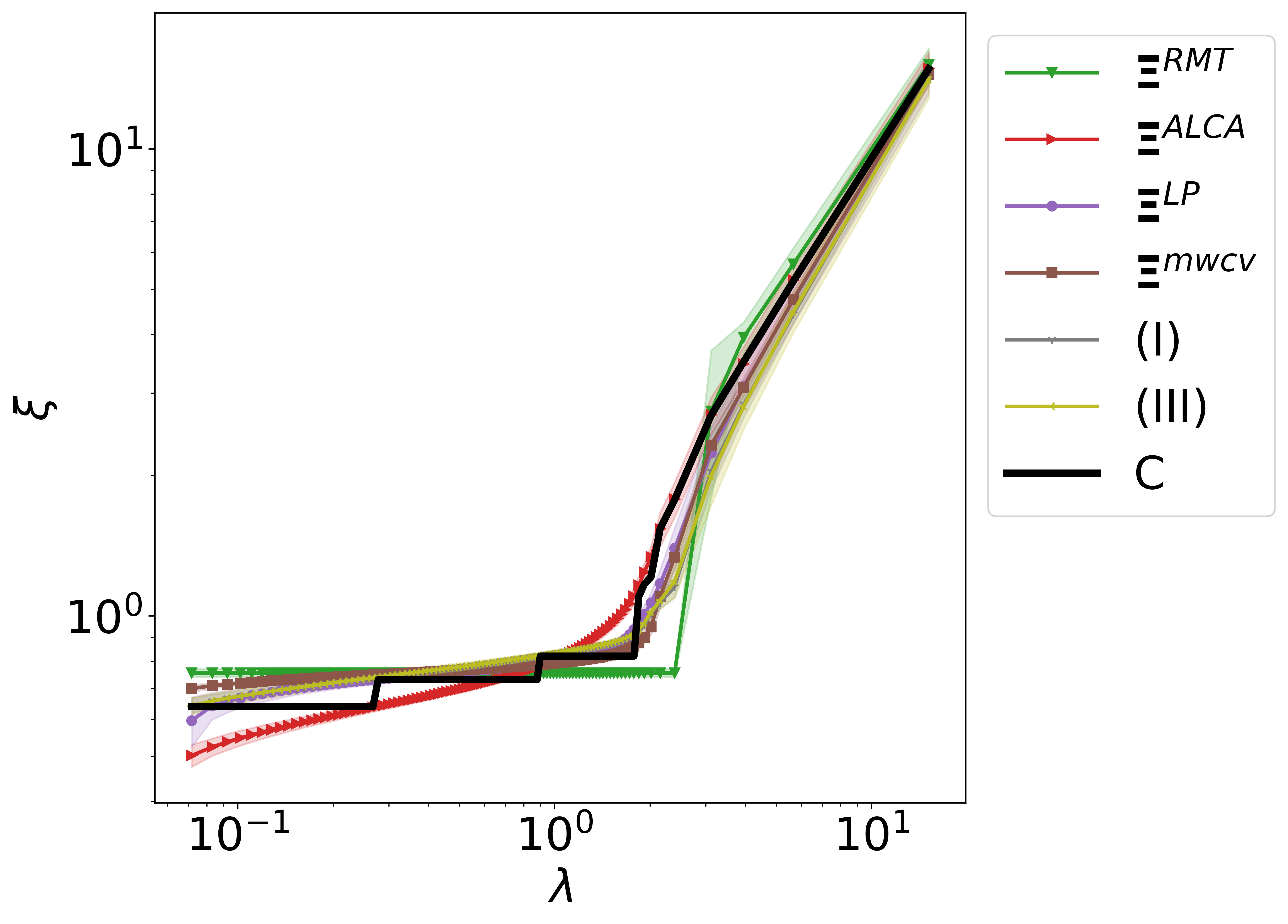}
         \caption{}
     \end{subfigure}
     \begin{subfigure}[b]{0.45\textwidth}
         \centering
        \includegraphics[scale=.3]{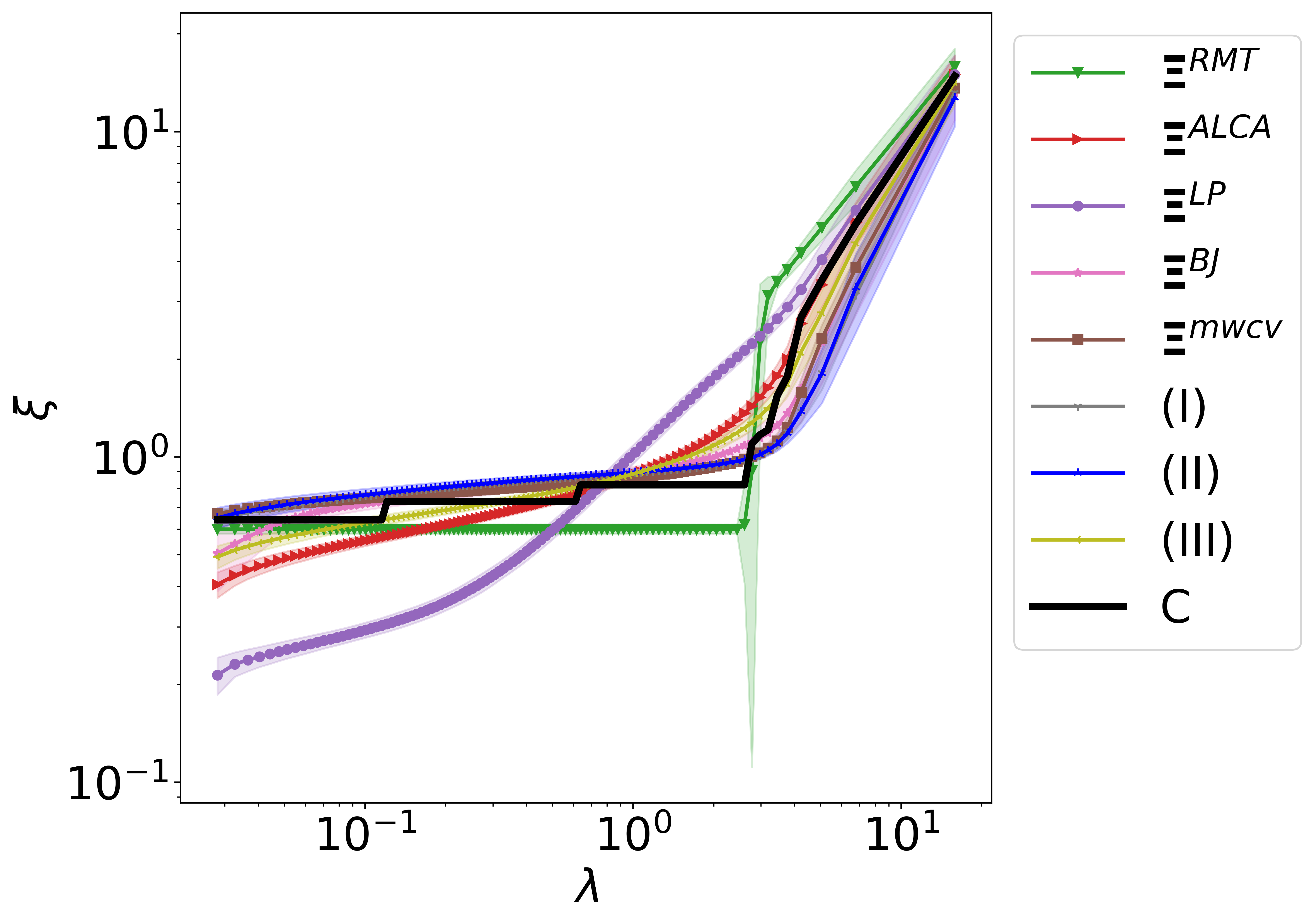}
         \caption{}
     \end{subfigure}
        \caption{Average shrinkage eigenvalues ($\xi$) vs.  average empirical eigenvalues ($\lambda$). (a) case 1. (b) case 2. (c) case 3.  The black line represents the corresponding population model $C$. The shadow band represents one standard deviation. Both axes are on logarithmic scales.}
        \label{Fig5}
\end{figure}

Figure~\ref{Fig5} shows the average shrinkage eigenvalues ($\xi$) as a function of the average empirical eigenvalues ($\lambda$) for case 1 (a), case 2 (b), and case 3 (c) under each of the considered filters. We can see the single-step filters do not deal well with the extreme eigenvalues, and the 2-step estimators somehow regularize the estimations. Notably,  the misspecification of the $\Xi^{LP}$ filter on case 3 presents a huge bias on almost the entire spectrum.
In general, the bias of the smallest eigenvalue has severe consequences on the metrics that require inverting the correlation matrix because a near singular matrix could be obtained. Hence the importance of correctly estimating these eigenvalues.

\begin{figure}[!ht]
     \begin{subfigure}[b]{0.45\textwidth}
         \centering
        \includegraphics[scale=.3]{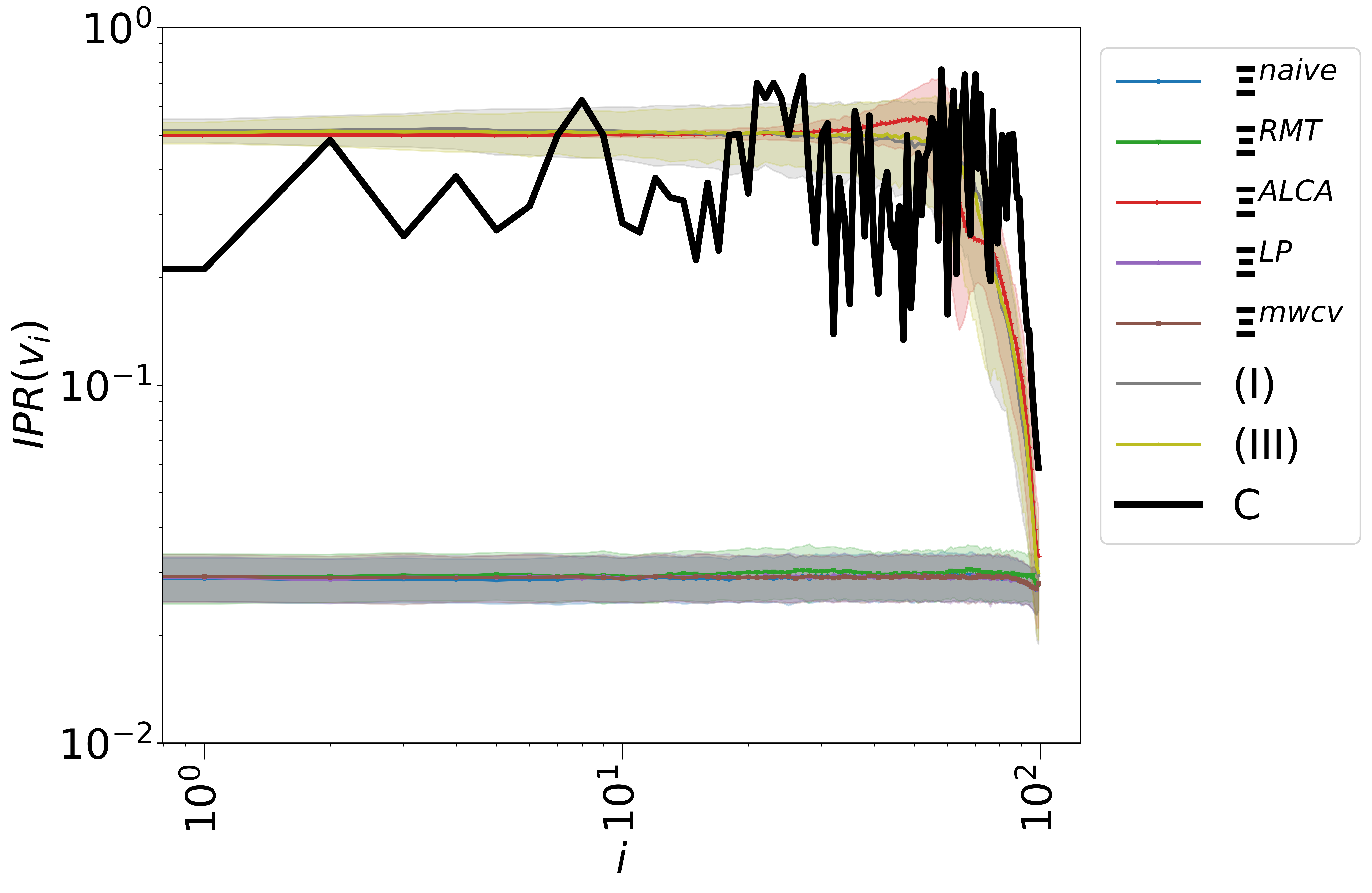}
         \caption{}
     \end{subfigure}
    \begin{subfigure}[b]{0.45\textwidth}
         \centering
        \includegraphics[scale=.3]{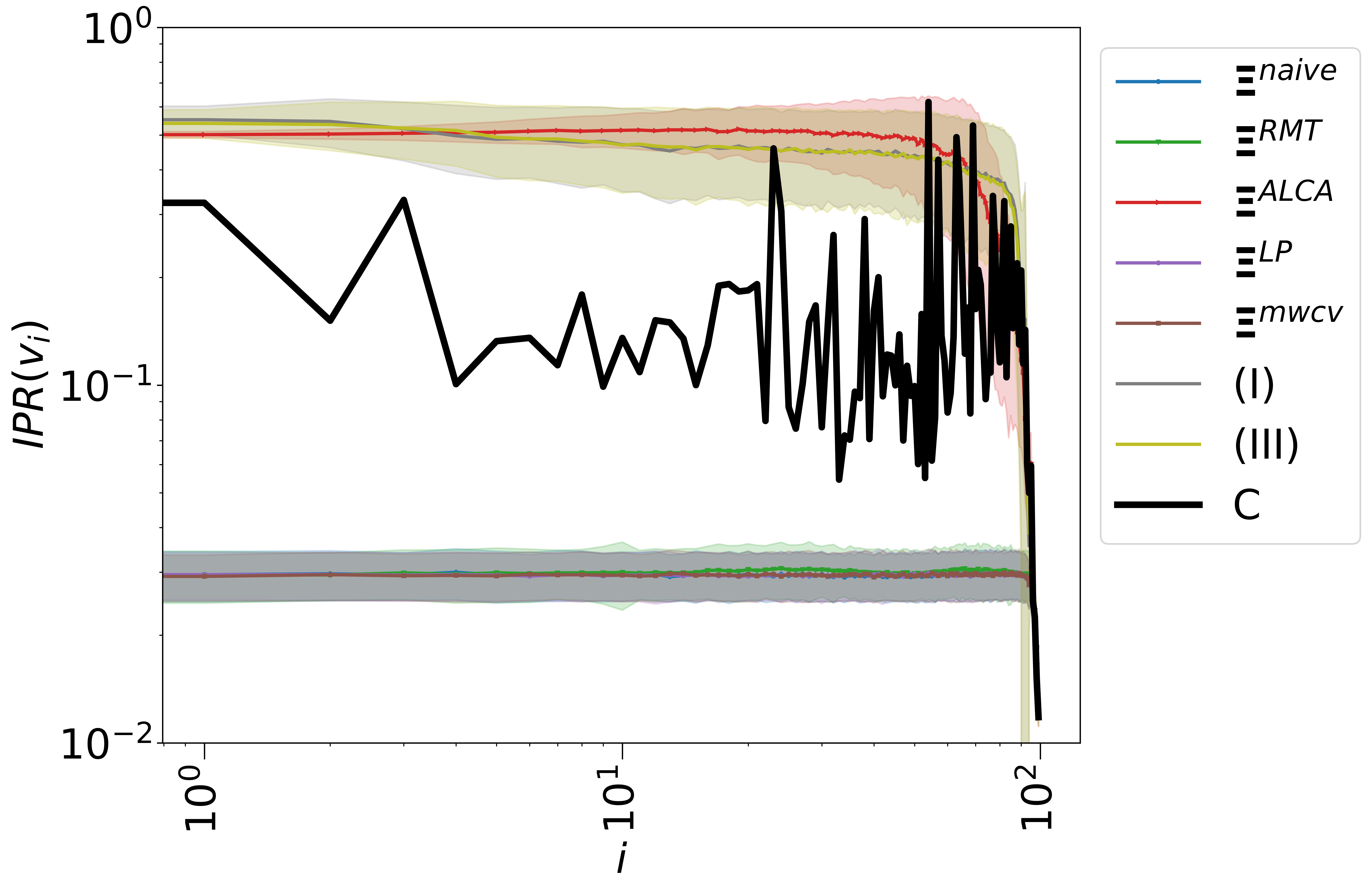}
         \caption{}
     \end{subfigure}
     \begin{subfigure}[b]{0.45\textwidth}
         \centering
        \includegraphics[scale=.3]{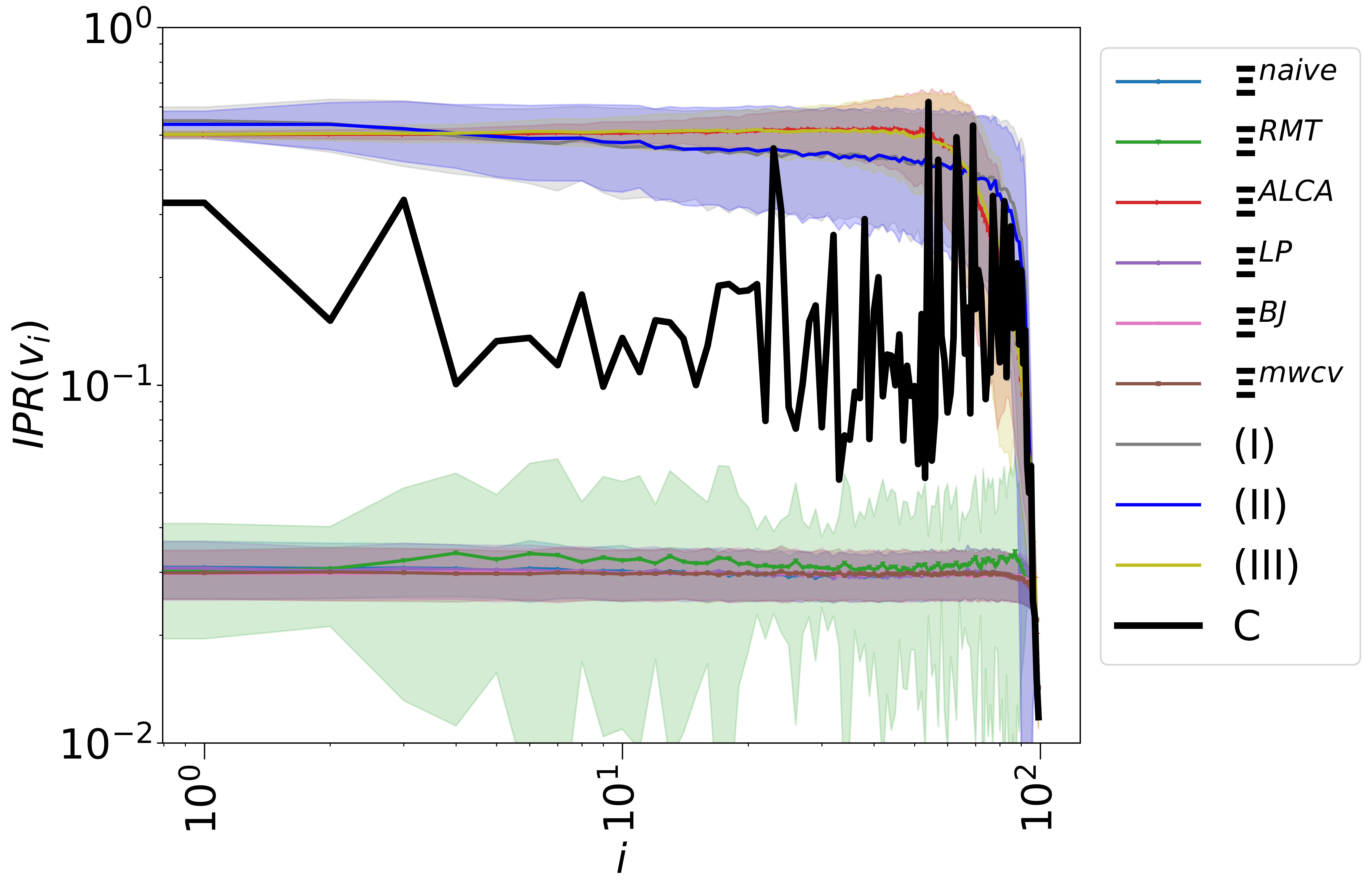}
         \caption{}
     \end{subfigure}
        \caption{Average IPR as a function of the rank of the $i$-th eigenvector of the filtered correlation matrix. (a) case 1. (b) case 2. (c) case 3.  The black line represents the corresponding population model $C$. The shadow band represents one standard deviation. Both axes are on logarithmic scales. The rank of eigenvectors runs from 1 (smallest eigenvalue) to 100 (largest eigenvalue).}
        \label{Fig6}
\end{figure}

On the other hand, the behavior of the eigenvectors can be characterized by the Inverse Participation Ratio~(IPR)~\cite{visscher1972localization}.
The IPR of the eigenvector $v_i$ is defined as~\cite{plerou2002random}: $IPR(v_i)=\sum_{j=1}^p [v_{i}^{(j)}]^4$; where $v_{i}^{(j)}$ is the $j$-th element of the eigenvector $v_i$. 
An eigenvector $v_i$  located in only one component has the upper bound $IPR(v_i) = 1$, while an eigenvector uniformly distributed over the $p$ components has the lower bound $IPR(v_i) = 1/p$.
Figure~\ref{Fig6} shows the average IPR as a function of the $i$-th eigenvector of the filtered correlation matrix  for case 1 (a), case 2 (b), and case 3 (c) under each of the considered filters.
It can be observed that the IPR of the eigenvectors related to the filters that fall into the RIE family present a stable behavior with a uniform distribution of its elements, which is natural since these filters assume that the eigenvectors do not change. The small fluctuations are due to the normalization effect to obtain orthonormal eigenvectors after reconstructing the correlation matrix.
On the contrary, the behavior of the eigenvectors that involve the ALCA filter is more localized and is closer to the eigenvectors of the population model~(black line).

\section{Discussion}

In principle, one would expect that asymptotic estimators based on random matrices and free probability perform better as the dimension of the correlation matrix increases. However, we have found at least one case where the hierarchical estimators perform better than the RIE estimators, even at $ p=500$~\cite{supplementary}. This behavior can be explained due to the assumptions of the semi-analytical solution of the non-linear shrinkage function proposed by Ledoit and Wolf. Their expression was also used in the Burda and Jarosz approach and our analysis. The central assumption of the solution is the existence of a compact interval that contains all the eigenvalues as the matrix dimensions tend to infinity. In other words, the eigenvalues should not grow with the dimension to converge to a well-defined density.
However, diagonal block and hierarchical nested models have one eigenvalue that grows with the dimension of each of their blocks~(see appendix~\ref{apendice1}). That is, a model of $k$ blocks has $k$ unbounded eigenvalues, which violates the assumptions of the asymptotic results of the RIE solutions. More precisely, the expressions in Eqs.~\ref{LedoitPeche2011}, \ref{BurdaJarosz2022} are correct and valid as long as $p,n\rightarrow \infty$. What is problematic is the kernel approximation of the density $\rho_E$ and the Hilbert transform $h_E$ since they do not converge for our block structure models. Hence, our models violate this principle. Thereupon,  the hierarchical clustering and 2-step estimators can give better estimates regarding optimality and stability.

Analyzing eigenvalues and eigenvectors reveals why the selected 2-step filters perform well against most metrics. If we change the order in constructing these estimators, we lose the regularizing effect by filtering the high-dimensional noise before applying the clustering methods. We have verified that estimating the smallest eigenvalues get worse by inverting the order of the single estimators composing the 2-step filters. This behavior is particularly noticeable for the inverse KL and inverse Frobenius metrics since they can be written as a function of the inverse of the eigenvalues. Therefore, if these are very close to zero, we obtain metrics that tend to infinity or indeterminate. On the part of the eigenvectors, the models that involve the hierarchical estimators modify the distribution of their elements and bring them closer to the population behavior qualitatively.
Nevertheless, block models present eigenvalues' multiplicity, and the set of eigenvectors is not unique. There may be slightly different solutions depending on the algorithm to compute them. Thus, a future question to explore is to what extent it is possible to filter the correlation matrix by modifying only the eigenvalues.

On the other hand, the excellent performance of the data-driven estimator $\mathbf{\Xi}^{mwcv}$ is notable. 
We have seen that the $\Xi^{mwcv}$ filter outperforms the two-step filters in case 3 under some metrics. 
A preliminary explanation is that the $\Xi^{mwcv}$ filter can capture autocorrelations due to its construction as a time-varying estimator. Then, the correlation matrix of non-stationary financial time series might be better estimated by the $\Xi^{mwcv}$ and 2-step (I) filters.
Furthermore, the authors of~\cite{bun2018overlaps} proved that it is possible to approximate the optimal RIE estimator $\xi(\lambda) = l$~(the true eigenvalues) by overlapping the eigenvectors of two different realizations of the same population covariance matrix $\Sigma$. Even valid if the test sample covariance matrix can be rank-deficient, i.e., $n=T_{out} < p$. Intuitively, the superposition of the training and testing eigenvectors helps estimate the empirical eigenvalues. As if rotating them into the test direction unveils their true value. This evidence opens the door to considering other types of non-linear shrinkage $\xi(\lambda)$ under the RIE approach. 

An interesting future work would consider statistical learning models to shape the function $\xi(\lambda)$ and consider the stylized fact of heterogeneous structures in financial correlation matrices under more general distributional assumptions.
Moreover, the non-linear shrinkage functions $\mathbf{\Xi}^{LP}$ and $\Xi^{BJ}$ are optimal concerning the Frobenius loss function.
Then, further future work also goes in the direction of analyzing the performance of the block diagonal and hierarchical nested models under a  non-linear shrinkage formula optimized having as a target the loss function used to evaluate their performance and in the spirit of the proposed expressions in~\cite{ledoit2018optimal,ledoit2021shrinkage, ledoit2022power}.

\begin{acknowledgments}
S.M. and R.N.M. acknowledge financial support by the MIUR PRIN project 2017WZFTZP, Stochastic forecasting in complex systems. A.G.M. acknowledge financial support by Consejo Nacional de Ciencia y Tecnolog\'ia (CONACYT, Mexico) through fund FOSEC SEP-INVESTIGACION BASICA (Grant No. A1-S-43514).
\end{acknowledgments}

\appendix

\section{}
\label{apendice2}
Tables~\ref{table4}, \ref{table5}, and \ref{table6} show the average stability  $\langle\mathcal{L}(\mathbf{\Xi}_i,\mathbf{\Xi}_j)\rangle$ of each estimator $\mathbf{\Xi}$ in relation to the loss function $\mathcal{L}$~(see tables~\ref{table1}, \ref{table2}, and \ref{table3} in the main text.)

\begin{table*}[!ht]
\small
\caption{Block diagonal model without memory (case 1). Stability of estimators in terms of $\langle \mathcal{L}(\mathbf{\Xi}_i,\mathbf{\Xi}_j)\rangle$, where $\mathcal{L}$ denotes the loss function and $\langle \cdot\rangle$ represents the average over the realizations of $m=1000$ and considering dimensions $p=100, n=200$.}
\begin{tabular}{|l|r|r|r|r|r|r|}
\hline
 & \multicolumn{1}{l|}{$\langle K(\mathbf{\Xi}_i,\mathbf{\Xi}_j)\rangle$} & \multicolumn{1}{l|}{$\langle K(\mathbf{\Xi}^{-1}_i,\mathbf{\Xi}^{-1}_j)\rangle$} & \multicolumn{1}{l|}{$\langle F(\mathbf{\Xi}_i,\mathbf{\Xi}_j)\rangle$} & \multicolumn{1}{l|}{$\langle F(,\mathbf{\Xi}^{-1}_i,\mathbf{\Xi}^{-1}_j)\rangle$} & \multicolumn{1}{l|}{$\langle MV(\mathbf{\Xi}_i,\mathbf{\Xi}_j)\rangle$} & \multicolumn{1}{l|}{$\langle SS(\mathbf{\Xi}_i,\mathbf{\Xi}_j)\rangle$} \\ \hline
$\mathbf{\Xi}^{naive}$ & 1.011480 & 1.011512 & 0.993376 & 9.102604 & 1.469460 & 2.022991 \\
$\mathbf{\Xi}^{RMT}$ & 0.030454 & 0.030248 & 0.185746 & 0.020132 & 0.038568 & 0.060702 \\
$\mathbf{\Xi}^{ALCA}$ & 0.050803 & 0.050802 & 0.147216 & 0.081805 & 0.082820 & 0.101605 \\
$\mathbf{\Xi}^{LP}$ & 0.015637 & 0.015614 & 0.042401 & 0.025937 & 0.027382 & 0.031251 \\
$\mathbf{\Xi}^{mwcv}$ & 0.010394 & 0.010385 & 0.030306 & 0.015103 & 0.017561 & 0.020779 \\
2-step (I) & {\bf 0.003341} & {\bf 0.003339} & {\bf 0.009333} & {\bf 0.004962} & {\bf 0.005754} & {\bf 0.006680} \\
2-step (III) & 0.004123 & 0.004108 & 0.011924 & 0.006024 & 0.007032 & 0.008231 \\
\hline
\end{tabular}
\label{table4}
\end{table*}

\begin{table*}[!ht]
\small
\caption{Hierarchical nested model without memory (case 2). Stability of estimators in terms of $\langle \mathcal{L}(\mathbf{\Xi}_i,\mathbf{\Xi}_j)\rangle$, where $\mathcal{L}$ denotes the loss function and $\langle \cdot\rangle$ represents the average over the realizations of $m=1000$ and considering dimensions $p=100, n=200$.}
\begin{tabular}{|l|r|r|r|r|r|r|}
\hline
 & \multicolumn{1}{l|}{$\langle K(\mathbf{\Xi}_i,\mathbf{\Xi}_j)\rangle$} & \multicolumn{1}{l|}{$\langle K(\mathbf{\Xi}^{-1}_i,\mathbf{\Xi}^{-1}_j)\rangle$} & \multicolumn{1}{l|}{$\langle F(\mathbf{\Xi}_i,\mathbf{\Xi}_j)\rangle$} & \multicolumn{1}{l|}{$\langle F(,\mathbf{\Xi}^{-1}_i,\mathbf{\Xi}^{-1}_j)\rangle$} & \multicolumn{1}{l|}{$\langle MV(\mathbf{\Xi}_i,\mathbf{\Xi}_j)\rangle$} & \multicolumn{1}{l|}{$\langle SS(\mathbf{\Xi}_i,\mathbf{\Xi}_j)\rangle$} \\ \hline
$\mathbf{\Xi}^{naive}$ & 1.015298 & 1.013145 & 0.952442 & 14.119574 & 1.188050 & 2.028444 \\
$\mathbf{\Xi}^{RMT}$ & 0.039437 & 0.039263 & 0.441274 & 0.022705 & 0.036337 & 0.078700 \\
$\mathbf{\Xi}^{ALCA}$ & 0.037348 & 0.037287 & 0.158198 & 0.096944 & 0.048786 & 0.074635 \\
$\mathbf{\Xi}^{LP}$ & 0.038438 & 0.038397 & 0.348333 & 0.051018 & 0.043579 & 0.076836 \\
$\mathbf{\Xi}^{mwcv}$ & 0.032339 & 0.032347 & 0.313248 & 0.030515 & 0.034258 & 0.064685 \\
2-step (I) & {\bf 0.012385} & {\bf 0.012390} & {\bf 0.084258} & {\bf 0.021132} & {\bf 0.015412} & {\bf 0.024775} \\
2-step (III) & 0.013008 & 0.012994 & 0.115886 & 0.022353 & 0.015728 & 0.026002 \\
\hline
\end{tabular}
\label{table5}
\end{table*}

\begin{table*}[!ht]
\small
\caption{Hierarchical nested model with memory (case 3). Stability of estimators in terms of $\langle \mathcal{L}(\mathbf{\Xi}_i,\mathbf{\Xi}_j)\rangle$, where $\mathcal{L}$ denotes the loss function and $\langle \cdot\rangle$ represents the average over the realizations of $m=1000$ and considering dimensions $p=100, n=200$.}
\begin{tabular}{|l|r|r|r|r|r|r|}
\hline
 & \multicolumn{1}{l|}{$\langle K(\mathbf{\Xi}_i,\mathbf{\Xi}_j)\rangle$} & \multicolumn{1}{l|}{$\langle K(\mathbf{\Xi}^{-1}_i,\mathbf{\Xi}^{-1}_j)\rangle$} & \multicolumn{1}{l|}{$\langle F(\mathbf{\Xi}_i,\mathbf{\Xi}_j)\rangle$} & \multicolumn{1}{l|}{$\langle F(,\mathbf{\Xi}^{-1}_i,\mathbf{\Xi}^{-1}_j)\rangle$} & \multicolumn{1}{l|}{$\langle MV(\mathbf{\Xi}_i,\mathbf{\Xi}_j)\rangle$} & \multicolumn{1}{l|}{$\langle SS(\mathbf{\Xi}_i,\mathbf{\Xi}_j)\rangle$} \\ \hline
$\mathbf{\Xi}^{naive}$ & 3.377219 & 3.384450 & 2.854999 & 99.721840 & 1.788246 & 6.761669 \\
$\mathbf{\Xi}^{RMT}$ & 0.288290 & 0.288862 & 2.137602 & 0.220317 & 0.223823 & 0.577152 \\
$\mathbf{\Xi}^{ALCA}$ & 0.156065 & 0.156478 & 0.760417 & 0.351341 & 0.181564 & 0.312543 \\
$\mathbf{\Xi}^{LP}$ & 0.723069 & 0.724795 & 1.899020 & 2.867135 & 0.610633 & 1.447864 \\
$\mathbf{\Xi}^{BJ}$ & 0.083587 & 0.084194 & 0.775484 & 0.534801 & 0.137637 & 0.167781 \\
$\mathbf{\Xi}^{mwcv}$ & 0.058353 & 0.058390 & 0.686160 & 0.056490 & 0.064800 & 0.116743 \\
2-step (I) & {\bf 0.039335} & {\bf 0.039237} & {\bf 0.438580} & {\bf 0.046152} & 0.046753 & {\bf 0.078572} \\
2-step (II) & 0.039901 & 0.039489 & 0.509946 & 0.047698 & {\bf 0.046319} & 0.079389 \\
2-step (III) & 0.096686 & 0.096735 & 0.615370 & 0.186985 & 0.118327 & 0.193421 \\
\hline
\end{tabular}
\label{table6}
\end{table*}

\section{}
\label{apendice1}

\subsection{Top eigenvalues of diagonal block and hierarchical nested models}
\label{derivation}

\subsubsection{Diagonal block model}
Consider a block diagonal matrix $\mathbf{A}$ of dimension $p\times p$  with $b$ blocks $\mathbf{A}_l (l=1,\dots, b)$ each of dimensions $p_l\times p_l$ satisfying $\sum_l p_l = p$
        \begin{equation}
        \mathbf{A} = \begin{pmatrix}
        \mathbf{A}_1 & 0 & \dots & 0 \\
        0 & \mathbf{A}_2 & \dots & 0 \\
        \dots  & \dots  & \dots & \dots \\
        0 & 0 & 0 & \mathbf{A}_b \end{pmatrix}
        \end{equation}
Be each block $\mathbf{A}_l$ of the form
        \begin{equation}
        \mathbf{A}_l = \begin{pmatrix}
        1 & a^{(l)} & \dots & a^{(l)} \\
        a^{(l)} & 1 & \dots & a^{(l)} \\
        \dots  & \dots  & \dots & \dots \\
        a^{(l)} & a^{(l)} & a^{(l)} & 1 \end{pmatrix},
        \end{equation}
where $a^{(l)}\in [0,1]$. The characteristic polynomial of $\mathbf{A}_l$ is found to be $\det(\mathbf{A}_l-\lambda\mathbf{I}) = (1-a^{(l)}-\lambda)^{p_l-1}(1+(p_l-1)a^{(l)}-\lambda)=0$. Thus, the eigenvalues of the block $\mathbf{A}_l$ are given by
         \begin{equation}
         \label{eigenvalues}
        \lambda = \left\lbrace\begin{array}{c}
        1 + a^{(l)}(p_l-1);  \quad \text{with multiplicity 1}\\
        \lambda = 1 -a^{(l)};  \quad \text{with multiplicity $p_l-1$}
        \end{array}\right.
         \end{equation}
The eigenvalues of $\mathbf{A}$ are the combined eigenvalues of each block due to the property
        \begin{equation}
        \det(\mathbf{A}-\lambda \mathbf{I}) = \det(\mathbf{A}_1 - \lambda \mathbf{I}) \dots \det(\mathbf{A}_b - \lambda \mathbf{I})
        \end{equation}
Therefore, there are $b$ eigenvalues of $\mathbf{A}$ that grow with the dimension of their blocks at the rate $p_l$. Consequently,  $b$ eigenvalues are not bounded when $p\rightarrow \infty$.

In addition, we can notice that $\mathbf{A}$ is reducible because there does not exist a directed path between the blocks in the associated directed graph $G(\mathbf{A})$, that is, $G(\mathbf{A})$ is not strongly connected~\cite{varga2000}. Nevertheless,  each directed subgraph $G(\mathbf{A}_l)$ is strongly connected given that $\mathbf{A}_l>O$. Then, each block matrix $\mathbf{A}_l$ is irreducible, and either
    \begin{equation}
        \sum_{j=1}^{p_l} [a_{ij}]_l = \rho(\mathbf{A}_l)\quad \text{for all}\quad 1\leq i \leq p_l,
    \label{bound1}
    \end{equation}
    or
    \begin{equation}
        \min_{1\leq i \leq p_l} \left( \sum_{j=1}^{p_l} [a_{ij}]_l  \right) < \rho(\mathbf{A}_l) < \max_{1\leq i \leq p_l} \left( \sum_{j=1}^{p_l} [a_{ij}]_l  \right),
    \label{bound2}
    \end{equation}
where $[a_{ij}]_l$ are the $(i,j)$-th elements of  $\mathbf{A}_l$, and $\rho(\mathbf{A}_l)$ is its \emph{spectral radius}.  
Further, the sum of each row of $\mathbf{A}_l$ is the same, then the minimum and maximum is equal. Hence we have
\begin{equation}
\label{radius}
\rho(\mathbf{A}_l) = \sum_{j=1}^{p_l} [a_{ij}]_l = 1 + (p_l-1)a^{(l)},
\end{equation}
Moreover, the generalization of the Perron-Frobenius theorem assures that $\mathbf{A}$ has a nonnegative real eigenvalue equal to its spectral radius. Therefore one of the spectral radius $\rho(\mathbf{A}_l), l=1,\dots, b$, is the spectral radius of $\mathbf{A}$. It can be corroborated that Eq. \ref{radius} coincides with the first part of Eq.~\ref{eigenvalues}.

\subsubsection{Hierarchical nested model}
We have something similar for the hierarchical nested model. In this case, the number of independent blocks is reduced. However, each of them is irreducible by the same argument given above. Let $\mathbf{A}_k$ be an independent hierarchical nested block, where $\sum_{k=1}^c p_k = p~(k=1,\dots, c)$, such that $c<b$. In other words, each independent hierarchical block is composed of several overlapping blocks. We have by construction
\begin{eqnarray}
 \min_{1\leq i \leq n} \left( \sum_{j=1}^n [a_{ij}]_k  \right) &=& 1 + (p_k-1)a^{(k)},\\
  \max_{1\leq i \leq n} \left( \sum_{j=1}^n [a_{ij}]_k   \right) &=& 1 + p_k(p_k-1)a^{(k)}.
\end{eqnarray}
The minimum is reached when no overlapping exists, and the model is reduced to the diagonal block model. The maximum is reached when each internal block overlaps with each other, and as we can have at most $p_k$ blocks, a factor $p_k$ appears. Then, Eqs.~\ref{bound1}-\ref{bound2} apply, and the bounds for the spectral radius of each independent block are
\begin{equation}
    1+(p_k-1)a^{(k)} < \rho(A_k) < 1+p_k(p_k-1)a^{(k)}
\end{equation}
Again, the generalization of the Perron-Frobenius theorem assures that $\mathbf{A}$ has a nonnegative real eigenvalue equal to its spectral radius. Thus, one of the spectral radius $\rho(\mathbf{A}_k)$, $k=1,\dots, c $, is the spectral radius of $\mathbf{A}$. Therefore, $c$ eigenvalues of $\mathbf{A}$ grow with the dimension of their blocks at the rate $p_k$~(at least). Consequently,  $c$ eigenvalues are not bounded when $p\rightarrow \infty$. 

\subsubsection{Observations}
We observe that the number of independent blocks in the hierarchical nested model  is less than the number in the diagonal block model, i.e., $c<b$. Consequently, the block's size of the former should be bigger to satisfy $\sum_{k=1}^c p_k = \sum_{l=1}^b p_l = p$. Therefore,  $p_k> p_l$, and  the top eigenvalue of the hierarchical nested model grows faster than the top eigenvalue of the diagonal block model.

In our models $a^{(k)} = a^{(l)} = \gamma^2 = (0.3)^2 = 0.09 $, the diagonal block model has $b=12$ diagonal blocks, while the hierarchical nested model has $c=3$ independent (non-overlapping) blocks. Then, it is clear that the top eigenvalue of the latter grows faster to infinity than the former as $p\rightarrow\infty$.

%

\end{document}